\documentclass[
    final,
    journal,
    letterpaper,
    twoside,
    twocolumn,
]{./style/IEEEtran}

\usepackage{subfigure}
\usepackage{amsmath}
\usepackage{amssymb}

\usepackage{cite}
\usepackage{graphicx}
\usepackage{array}
\usepackage{graphicx}

\usepackage{balance}   
\usepackage{color}
\usepackage[colorlinks=false, linkcolor=blue, bookmarks=false]{hyperref}

\newtheorem{proposition}{\textbf{Proposition}}

\newtheorem{remark}{\textbf{Remark}}

\newcommand{\Sumin}[1]{\textcolor{black}{{#1}}}

\def\SR{\mathop{\{\mathcal{S}\hspace{-0.09cm} \rightarrow \hspace{-0.09cm} \mathcal{R}\}}}
\def\RD{\mathop{\{\mathcal{R}\hspace{-0.09cm} \rightarrow \hspace{-0.09cm} \mathcal{D}\}}}
\def\RR{\mathop{\{\mathcal{R}\hspace{-0.09cm} \rightarrow \hspace{-0.09cm} \mathcal{R}\}}}

\def\Si{\mathop{\{\mathcal{S}\hspace{-0.09cm} \rightarrow \hspace{-0.09cm} i\}}}
\def\jD{\mathop{\{j \hspace{-0.09cm} \rightarrow \hspace{-0.09cm} \mathcal{D}\}}}
\def\ji{\mathop{\{j \hspace{-0.09cm} \rightarrow \hspace{-0.09cm} i\}}}

\begin{document}
\title{Virtual Full-Duplex Buffer-Aided Relaying in the Presence of Inter-Relay Interference
\author{Su~Min~Kim,~\IEEEmembership{Member,~IEEE} and Mats~Bengtsson,~\IEEEmembership{Senior Member,~IEEE}}
\thanks{S. M. Kim was with the School of Electrical Engineering, KTH Royal Institute of Technology and is currently with the Department of Electronics Engineering, Korea Polytechnic University, Siheung, Korea (E-mail: suminkim@kpu.ac.kr).} 
\thanks{M. Bengtsson is with the School of Electrical Engineering, KTH Royal Institute of Technology, Stockholm, Sweden (E-mail: mats.bengtsson@ee.kth.se).}
\thanks{Part of this work has been presented at \emph{IEEE International Symposium on Personal, Indoor, Mobile and Radio Communications (PIMRC)}, London, UK, 2013.}
\thanks{This work has been performed in the framework of the FP7 project ICT-317669 METIS. The authors would like to acknowledge the contributions of their colleagues. This information reflects the consortium’s view, but the consortium is not liable for any use that may be made of any of the information contained therein.}
\thanks{This work has been submitted to the IEEE for possible publication. Copyright may be transferred without notice, after which this version may no longer be accessible.}
}


\maketitle

\begin{abstract}
In this paper, we study virtual full-duplex (FD) buffer-aided relaying to recover the loss of multiplexing gain caused by half-duplex (HD) relaying in a multiple relay network, where each relay is equipped with a buffer and multiple antennas, through joint opportunistic relay selection (RS) and beamforming (BF) design.
The main idea of virtual FD buffer-aided relaying is that the source and one of the relays simultaneously transmit their own information to another relay and the destination, respectively. 
In such networks, inter-relay interference (IRI) is a crucial problem which has to be resolved like self-interference in the FD relaying. 
In contrast to previous work that neglected IRI, we propose joint RS and BF schemes taking IRI into consideration by using multiple antennas at the relays.
In order to maximize average end-to-end rate, we propose a weighted sum-rate maximization strategy assuming that adaptive rate transmission is employed in both the source to relay and relay to destination links. Then, we propose several BF schemes cancelling or suppressing IRI in order to maximize the weighted sum-rate.
Numerical results show that our proposed optimal, zero-forcing, and minimum mean square error BF-based RS schemes asymptotically approach the ideal FD relaying upper bound when increasing the number of antennas and/or the number of relays.
\end{abstract}


\section{Introduction}

Since cooperative relaying can improve both spectral efficiency and spatial diversity, it is a promising core technology for next-generation wireless communication networks.
So far, most studies have considered half-duplex (HD) relaying based on two-phase operation where a source transmits data to relays at the first time slot and the relays forward it to a destination at the second time slot \cite{LTW04TIT,LW04TIT}. However, such HD relaying causes a \emph{loss of multiplexing gain} expressed as an one-half pre-log factor.
To overcome the loss of multiplexing gain, several practical full-duplex (FD) relaying solutions have been studied \cite{DS10Asilomar,CJS+10MC,JCK+11MC,DDS12TWC,AKS+12MC,BMK13SC,DSA+14TVT,BK14NSDI,BJK14SC,BK14SC}.
Since \emph{strong self-interference} is a main problem which has to be resolved in FD relaying, the previous work primarily focused on self-interference cancellation based on antenna separation techniques in the wireless propagation domain and signal cancellation techniques in the analog circuit and digital domains.
Although these studies showed feasibility of FD relaying using small-scale wireless communication devices such as WiFi and IEEE 802.15.4, the technology is still premature for cellular communications, which require additional cancellation gains due to practical limitations such as varying center frequencies, bandwidth, and circuit imperfections.

In order to mitigate the loss of multiplexing gain in HD relaying, \emph{successive relaying} protocols have been proposed for a two-relay network \cite{RW07JSAC,FWT+07TWC,RGK10TIT,MK10TC,WFT+10TWC} and multiple-relay networks \cite{TaN08JSAC,HLT12TWC,KCJ+13TWC}. 
In these protocols, two relays take turns acting as receivers and transmitters successively and a source and a transmitting relay transmit their own information simultaneously. Here, the source transmits new information and the relay transmits previously received information.
The main issue for such successive relaying protocols is to efficiently handle \emph{inter-relay interference} (IRI) from the transmitting relay to the receiving relay.
Towards this end, successive interference cancellation (SIC) and/or sophisticated coding and joint decoding techniques have been employed in the literature.
However, the SIC requires strong interference scenarios and the joint decoding requires high computational complexity.
Furthermore, although the successive relaying asymptotically achieves the spectral efficiency of the FD relaying with respect to the number of channel uses, it requires a sufficiently long block length (equivalently, coherence time) over slow fading channels.

Employing a buffer at the relay, such long block length constraints can be relaxed. Focusing on these advantages, \emph{buffer-aided relaying} has been proposed in a three-node network \cite{XFT+08TWC,ZSP11GC,ZlS13TIT,ZSP13JSAC}. The key idea is an opportunistic relaying mode selection (buffering or forwarding) according to channel conditions. 
HD buffer-aided relaying can achieve up to two-fold spectral efficiency under asymmetric channel conditions between $\SR$ and $\RD$ links, compared to HD relaying without buffer.
Additionally, bidirectional buffer-aided relaying with two-way traffic \cite{LPC+13CL,JZI+13EUSIPCO,JZI+13GC,SZA+13ISIT}, buffer-aided relaying over dual-hop broadcast channels \cite{ZSA+13TWC} and a shared relay channel with two source-destination pairs \cite{ZSA+13TC} have been studied.
By extending to multiple-relay networks, several \emph{opportunistic relaying} schemes, which exploit the best HD buffer-aided relay, have been proposed \cite{IMS12TWC,KCT12TWC,IKS12TVT}.
Ikhlef \emph{et al.} \cite{IMS12TWC} have proposed a $\max-\max$ relay selection (MMRS) scheme, which selects the best $\SR$ and $\RD$ relays with the maximum channel gains.
However, the MMRS scheme does not fully take advantage of the benefits of buffer-aided relaying since it maintains the two-phase operation.
Therefore, Krikidis \emph{et al.} \cite{KCT12TWC} have proposed a $\max-\mathrm{link}$ relay selection (MLRS) scheme, which selects the best relaying mode as well as the maximum channel gain. 

Most recently, Ikhlef \emph{et al.} \cite{IKS12TVT} have proposed a space full-duplex $\max-\max$ relay selection (SFD-MMRS) scheme, which mimics the FD relaying by utilizing the best receiving and transmitting relays operating simultaneously.
In this scheme, they did not consider IRI by assuming fixed-relays with highly directional antennas.
However, this assumption does not always hold and it is hard to be practically realized as the number of relays increases.
With consideration of IRI, Kim and Bengtsson \cite{KB13PIMRC} proposed a virtual FD buffer-aided relaying scheme based on opportunistic relay selection (RS) with zero-forcing beamforming (BF) for IRI cancellation in order to maximize average end-to-end rate assuming \emph{adaptive rate transmission}.
Nomikos \emph{et al.} \cite{NCK+13PIMRC,NVC+13TETT} have proposed a buffer-aided successive opportunistic relaying (BA-SOR) scheme employing SIC at the receiving relay for \emph{fixed rate transmission}. 
In \cite{NCK+13PIMRC}, even if it partially overcame the strong interference requirement of SIC through power allocation at the source and relays, the main objective was to minimize the total energy expenditure. 
In \cite{NVC+13TETT}, the average end-to-end rate of the BA-SOR scheme, which has been originally devised for fixed rate transmission, is numerically shown for adaptive rate transmission. However, a fixed low SIC threshold, $r_0$ (e.g., 2 bps/Hz), has been applied even for adaptive rate transmission whereas the threshold value should be set to the information rate of the $\RD$ link. 
Thus, this yields an optimistic result in terms of the average throughput. 

In this paper, our main goal is to approach the average end-to-end rate of ideal FD relaying even in the presence of IRI.
To this end, we propose transmission schemes based on a joint RS and BF design utilizing multiple buffer-aided relays and multiple antennas at the relays.
For the joint RS and BF design, we first propose a weighted sum-rate maximization using instantaneous channel and buffer states for achieving the average end-to-end rate maximization.
Then, we separately design linear BF for each (receiving and transmitting) relay pair, which cancels or suppresses IRI, and optimal RS for maximizing the weighted sum-rate based on the beamformers found for each relay pair.
To focus on maximizing the average end-to-end rate, we employ adaptive rate transmission at the source and relays (i.e., channel state information at transmitter for both nodes) and consider delay-tolerant applications.
Our main contributions in this work are summarized as follows:
\begin{itemize}
\item A new RS criterion based on a weighted sum of instantaneous rates is proposed to maximize the average end-to-end rate in a virtual FD buffer-aided relaying network with adaptive rate transmission.
\item Various transmit and receive BF design strategies at the multiple antenna relays are proposed in order to cancel or suppress IRI.
\item We show that joint RS and BF schemes achieve the ideal FD relaying bound in terms of the average end-to-end rate asymptotically with increasing the number of antennas and/or the number of relays.
\item Compared to our previous work \cite{KB13PIMRC}, we propose more practical joint RS and BF schemes which can support non-identical channel conditions, including an iterative optimal BF-based RS scheme which achieves the best average end-to-end rate. Moreover, we provide extensive numerical results including average end-to-end rate, average delay, effect of IRI intensity, behavior of optimal weight factors, and effect of finite buffer size.
\end{itemize}

The rest of this paper is organized as follows. In Section~\ref{sec:system_model}, the system model is presented. The instantaneous rates and average end-to-end rate of a buffer-aided relaying network are described in Section~\ref{sec:capacity}. Buffer-aided joint RS and BF schemes considering IRI are proposed in Section~\ref{sec:relay_selection}. 
In Section~\ref{sec:performance_evaluation}, the performance of the proposed schemes are evaluated through simulations. Finally, conclusive remarks and future work are provided in Section~\ref{sec:conclusion}.

\section{System Model} \label{sec:system_model}

\begin{figure}
\centering
\includegraphics[width=6cm]{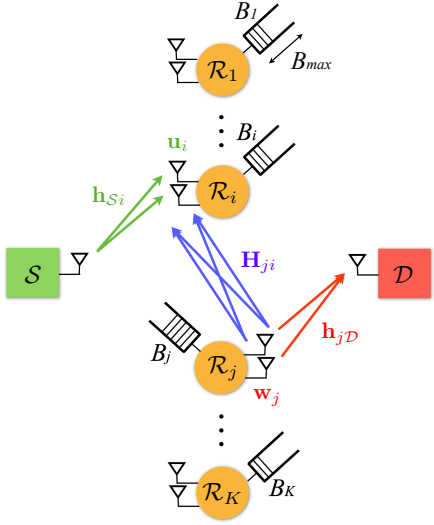}
\caption{System Model: a single source, a single destination, and multiple relays with buffer and multiple antennas (e.g., $M=2$).}
\label{fig:sys_model}
\vspace{-3mm}
\end{figure}

In this paper, we consider a source, $\mathcal{S}$, and a destination, $\mathcal{D}$, which have a single antenna, and $K$ buffer-aided relays with $M$ antennas each (e.g., in Fig.~{\ref{fig:sys_model}}, $M=2$).
Denote the set of HD buffer-aided decode-and-forward relays by $\mathcal{K}=\{1,\ldots,K\}$.
We assume that there is no direct path between the source and destination as in the related literature \cite{TaN08JSAC,HLT12TWC,KCJ+13TWC,XFT+08TWC,ZSP11GC,ZlS13TIT,ZSP13JSAC,IMS12TWC,KCT12TWC,IKS12TVT,KB13PIMRC,NCK+13PIMRC,NVC+13TETT}.
This system model can be regarded as an example of relay-assisted device-to-device communications where the source and destination are low-cost devices with some limitations such as a single antenna.
The source is supposed to always have data traffic to transmit.
In addition, let $\mathbf{h}_{\mathcal{S}i}$, $\mathbf{h}_{j\mathcal{D}}$, and $\mathbf{H}_{ji}$, $i,j\in\mathcal{K}$ denote the channel coefficient vectors and matrices of  $\Si$, $\jD$, and $\ji$ links, respectively.
The channel fading is assumed to be stationary and ergodic and in the asymptotic analysis and numerical examples, we will make the additional assumption that all the channel coefficients follow circular symmetric complex Gaussian distributions, $\mathbf{h}_{\mathcal{S}i}\sim\mathcal{CN}(0,\sigma_{\mathcal{S}\mathcal{R}_i}^2 \mathbf{I})$, $\mathbf{h}_{j\mathcal{D}}\sim\mathcal{CN}(0,\sigma_{\mathcal{R}_j\mathcal{D}}^2 \mathbf{I})$, and $\mathrm{vec}[{\mathbf{H}_{ji}}]\sim\mathcal{CN}(0,\sigma_{\mathcal{R}_j\mathcal{R}_i}^2 \mathbf{I})$ where $\mathrm{vec}[\cdot]$ denotes the vectorization of a matrix.

In order to mimic FD relaying, $\SR$ and $\RD$ transmissions are performed simultaneously by using the best pair of receiving and transmitting relays as in \cite{IKS12TVT,NCK+13PIMRC,NVC+13TETT,KB13PIMRC}.
To this end, the receiving relay decodes the data received from the source and stores it in its buffer, while the transmitting relay encodes data from its buffer and sends it to the destination.
For a given selected relay pair $(i,j)$, $i\neq j$, the received signal vector at the receiving relay $i$ is
\begin{align}
\mathbf{y}_i^{(i,j)} &= \mathbf{h}_{\mathcal{S}i} x_{\mathcal{S}} + \mathbf{H}_{ji} \mathbf{x}_{j} + \mathbf{n}_i = \mathbf{h}_{\mathcal{S}i} x_{\mathcal{S}} + \mathbf{H}_{ji} \mathbf{w}_{j} x_j + \mathbf{n}_i,
\label{eq:recv_sig_vec_sr}
\end{align} 
where 
$\mathbf{h}_{\mathcal{S}i}\in\mathbb{C}^{M \times 1}$ denotes the channel vector from the source to the $i$-th relay,
$\mathbf{H}_{ji}\in\mathbb{C}^{M \times M}$ denotes the inter-relay channel matrix from the $j$-th relay to the $i$-th relay, $x_{\mathcal{S}}$ denotes the transmitted data symbol from the source, and $\mathbf{x}_{j} = \mathbf{w}_{j} x_j$ denotes the transmitted data symbol vector from the $j$-th relay where $\mathbf{w}_{j}=[w_{j}^{1},\ldots,w_{j}^{M}]^T$ and $x_j$ represent the transmit BF vector of the $j$-th relay and the transmitted data symbol of the $j$-th relay, respectively. Here, $\mathbb{E}\left[|x_{\mathcal{S}}|^2\right] \leq {P}_{\mathcal{S}}$ and $\mathbb{E}\left[|x_j|^2\right] \leq {P}_{\mathcal{R}}$ where ${P}_{\mathcal{S}}$ and ${P}_{\mathcal{R}}$ denote the maximum transmit powers of the source and the relay, respectively, and $\| \mathbf{w}_{j} \| = 1$ for $i,j\in\mathcal{K}$ where $\|\cdot\|$ denotes the 2-norm. $\mathbf{n}_i$ denotes an additive white Gaussian noise (AWGN) vector with zero mean and covariance $\sigma_n^2\mathbf{I}$, i.e., $\mathbf{n}_i\sim\mathcal{CN}(0,\sigma_n^2\mathbf{I})$.
For a given selected relay pair $(i,j)$, $i \neq j$, the received signal at the destination is
\begin{align}
y_{\mathcal{D}}^{(i,j)} &= \mathbf{h}_{j\mathcal{D}}^H\mathbf{x}_{j} + n_{\mathcal{D}} 
= \mathbf{h}_{j\mathcal{D}}^H\mathbf{w}_{j} x_j + n_{\mathcal{D}},
\label{eq:recv_sig_rd}
\end{align}
where 
$\mathbf{h}_{j\mathcal{D}}\in\mathbb{C}^{M \times 1}$ denotes the channel vector from the $j$-th relay to the destination\footnote{For notational convenience, we define $\mathbf{h}_{j\mathcal{D}}$ as the complex conjugate channel vector differently from the definition of $\mathbf{h}_{\mathcal{S}i}$, i.e., $\mathbf{h}_{j\mathcal{D}}^H=[h_{j\mathcal{D}}^{1},\ldots,h_{j\mathcal{D}}^{M}]$ while $\mathbf{h}_{\mathcal{S}i}=[h_{\mathcal{S}i}^{1},\ldots,h_{\mathcal{S}i}^{M}]^T$.}, $(\cdot)^H$ denotes the Hermitian transpose,
and $n_{\mathcal{D}}$ denotes AWGN with zero mean and variance $\sigma_n^2$, i.e., $n_{\mathcal{D}}\sim\mathcal{CN}(0,\sigma_n^2)$.

\section{Instantaneous and Average Data Rates of a Buffer-Aided Relaying Network} \label{sec:capacity}

Using a linear receive BF vector $\mathbf{u}_{i}=[u_{i}^{1},\ldots,u_{i}^{M}]^T$ at the $i$-th receiving relay, with $\|\mathbf{u}_i\|=1$, the received signal after the receive BF becomes
\begin{align}
r_{i}^{(i,j)} = \mathbf{u}_i^H \mathbf{y}_{i}^{(i,j)}
&=\mathbf{u}_i^H \mathbf{h}_{\mathcal{S}i} x_{\mathcal{S}} + \mathbf{u}_i^H \mathbf{H}_{ji} \mathbf{w}_{j} x_j + \tilde{n}_i,
\label{eq:recv_sig_sr}
\end{align}
for given relay pair $(i,j)$, where $\tilde{n}_i= \mathbf{u}_i^H \mathbf{n}_i\sim \mathcal{CN}(0,\sigma_n^2)$.
From \eqref{eq:recv_sig_sr} and \eqref{eq:recv_sig_rd}, the instantaneous received SINR/SNR for the $\Si$ and $\jD$ links at time slot $t$ are expressed, respectively, as:
\begin{align}
\gamma_{\mathcal{S}i}^{(i,j)}(t) 
&= \frac{\rho_{\mathcal{S}} | \mathbf{u}_{i}^{H}\mathbf{h}_{\mathcal{S}i} |^2 }{ 1 + \rho_{\mathcal{R}} |\mathbf{u}_{i}^{H} \mathbf{H}_{ji}\mathbf{w}_{j} |^2  },\label{eq:SINRi}\\
\gamma_{j\mathcal{D}}^{(i,j)}(t) 
&= \rho_{\mathcal{R}} | \mathbf{h}_{j\mathcal{D}}^H \mathbf{w}_{j} |^2,
\label{eq:SNRD}
\end{align} 
where $\rho_{\mathcal{S}} = P_{\mathcal{S}} / \sigma_n^2$ and $\rho_{\mathcal{R}} = P_{\mathcal{R}} / \sigma_n^2$.
Let $B_i(t)$ denote the number of bits in the buffer normalized by the number of channel uses of the $i$-th relay at the end of time slot $t$. 
Assuming a Gaussian codebook and information theoretic capacity achieving coding scheme, and taking the buffer state at the receiving relay $i$ into consideration, the instantaneous rate of the $\Si$ link is
\begin{align}
C_{\mathcal{S}i}^{(i,j)}(t) = \min\left\{ \log_2\left(1+\gamma_{\mathcal{S}i}^{(i,j)}(t)\right), B_{max}-B_i(t-1) \right\},
\label{eq:C_si}
\end{align}
where $\min\{\cdot, \cdot\}$ denotes the minimum value of arguments, $B_{max}$ denotes the maximum buffer size, and the buffer of the $i$-th relay is updated by
\begin{align}
B_i(t) = B_i(t-1) + C_{\mathcal{S}i}^{(i,j)}(t).\nonumber
\end{align}
For the transmitting relay $j$, the instantaneous rate of link $\jD$ in time slot $t$ is
\begin{align}
C_{j\mathcal{D}}^{(i,j)}(t) = \min\left\{ \log_2\left(1+\gamma_{j\mathcal{D}}^{(i,j)}(t)\right), B_j(t-1) \right\},
\label{eq:C_jd}
\end{align}
where the buffer of the $j$-th relay is updated by
\begin{align}
B_j(t) = B_j(t-1) - C_{j\mathcal{D}}^{(i,j)}(t).\nonumber
\end{align}

The average received data rate at the destination, over a time window of length $T$, is given by
\begin{equation}
  \label{eq:dest_rate}
  \bar{C}_{\mathcal{D}} = \frac{1}{T} \sum_{t=1}^T C_{j(t)\mathcal{D}}^{(i(t),j(t))}(t)\;,
\end{equation}
where $i(t)$ and $j(t)$ denote the indices of the selected receiving and transmitting relay indices in time slot $t$, respectively. Similarly, the average transmitted data rate at the source is 
\begin{equation}
  \label{eq:source_rate}
  \bar{C}_{\mathcal{S}} = \frac{1}{T} \sum_{t=1}^T C_{\mathcal{S}i(t)}^{(i(t),j(t))}(t)\;.
\end{equation}

\section{Buffer-Aided Joint Relay Selection and Beamforming Schemes in the Presence of Inter-Relay Interference} \label{sec:relay_selection}

Our objective is to maximize the average data rate given in~\eqref{eq:dest_rate}, through a joint RS and BF design. Optimizing the minimum of the instantaneous rates, which corresponds to maximizing a lower bound on $\min\{\bar{C}_{\mathcal{S}},\bar{C}_{\mathcal{D}}\}$ \cite{XFT+08TWC}, was proposed in \cite{BKR+06JSAC} as an HD best relay selection criterion. Since this is suboptimal in the presence of IRI, we here propose an alternative approach, by formulating the following optimization problem, 
\begin{subequations}
    \label{eq:C_D_opt}
    \begin{IEEEeqnarray}{cl}
      \hspace{-8mm}\underset{\{\mathbf{u}_{i(t)},\mathbf{w}_{j(t)}, i(t), j(t)\}}{\mbox{max}}~ & \bar{C}_{\mathcal{D}} \\
      \hspace{-12mm}\mbox{s. t.} & \frac{1}{T} \sum_{t=1}^T \Big[ C_{\mathcal{S}k}^{(i(t),j(t))}(t)\nonumber\\
      \hspace{-12mm}& \hspace{8mm}- C_{k\mathcal{D}}^{(i(t),j(t))}(t) \Big] \geq 0, ~\forall k \in \mathcal{K}, \label{eq:non-decreasing_buffers} \\
      \hspace{-12mm}& \|\mathbf{u}_{i(t)}\| \leq 1, \|\mathbf{w}_{j(t)}\| \leq 1,\label{eq:BF_norms}\\
      \hspace{-12mm}& i(t) \neq j(t).\label{eq:constraint_neq}
  \end{IEEEeqnarray}
\end{subequations}
In order for the relay buffers to stay non-empty, we have here added the constraints~\eqref{eq:non-decreasing_buffers} (the buffer is absorbing). Obviously, we want~\eqref{eq:non-decreasing_buffers} to hold with equality (buffer stability), but the inequality formulation is exploited below to determine constrains on the weighting parameters. Problem~\eqref{eq:C_D_opt} is non-convex in the beamforming vectors and combinatorial in the relay selection. For tractability, we therefore use a Lagrange relaxation~\cite{Fisher:2004}, studying the partial Lagrangian with dual variables $\alpha_k$ corresponding to the constraints~\eqref{eq:non-decreasing_buffers}, 
\begin{align}
  \label{eq:Lagr_C_D}
  &\mathcal{L}(\mathbf{u}_{i(t)},\mathbf{w}_{j(t)}, i(t), j(t),\{\alpha_k\})\nonumber\\
  &= \bar{C}_{\mathcal{D}} + 
  \sum_{k\in\mathcal{K}} \alpha_k \left( \frac{1}{T} \sum_{t=1}^T \left[ C_{\mathcal{S}k}^{(i(t),j(t))}(t) - C_{k\mathcal{D}}^{(i(t),j(t))}(t) \right] \right). 
\end{align}
Similarly to~\cite{XFT+08TWC}, and for the moment assuming unbounded buffers, it can be shown that the probability that $B_j(t-1) < \log_2\left(1+\gamma_{j\mathcal{D}}^{(i,j)}(t)\right)$ goes to zero, if the channel fading is stationary and~\eqref{eq:non-decreasing_buffers} holds. Therefore, the RS and BF can be optimized separately for each time step. Collecting the terms corresponding to time step $t$, the primal variables optimizing $\mathcal{L}(\cdot)$ are given by
\begin{subequations}
  \label{eq:weighted_opt}
  \begin{IEEEeqnarray}{cl}  
    \underset{\{\mathbf{u}_{i},\mathbf{w}_{j}, i, j\}}{\mbox{max}}~ & \alpha_i C_{\mathcal{S}i}^{(i,j)} + (1-\alpha_j) C_{j\mathcal{D}}^{(i,j)} \\
    \mbox{s. t.} & \eqref{eq:BF_norms}\quad \&\quad \eqref{eq:constraint_neq}
  \end{IEEEeqnarray}
\end{subequations}
and it follows from the inequality in~\eqref{eq:non-decreasing_buffers} that $\alpha_k\geq 0$ at the optimum. A similar argument, instead optimizing the average transmit rate at the source, under constraints that the average relay buffers are non-increasing, with Lagrange multipliers $1-\alpha_k$, gives exactly the same criterion~\eqref{eq:weighted_opt}, but with the constraints that $1-\alpha_k\geq 0$. The conclusion of this Lagrange relaxation is therefore that the optimal RS and BF strategy can be done separately for each time step based on instantaneous information and has the form~\eqref{eq:weighted_opt}, for some choice of weight factors (dual variables) $\alpha_k$ with $0\leq \alpha_k\leq 1$. 

One possible approach to determine the optimal $\alpha_k$ is to use subgradient optimization. As shown in~\cite{Bertsekas:95}, a subgradient for $\alpha_k$ is given by the left hand side of~\eqref{eq:non-decreasing_buffers}. However, to avoid having to run the system $T$ time steps between each update of $\alpha_k$, we propose to replace the time average in~\eqref{eq:non-decreasing_buffers} by an exponentially weighted sum
  \begin{equation}
    \label{eq:exp_aver_queue_change}
    \Delta_{B_k}(t) = (1-\lambda) \sum_{\tau=0}^t  \lambda^{t-\tau} \left[ C_{\mathcal{S}k}^{(i(\tau),j(\tau))}(\tau) - C_{k\mathcal{D}}^{(i(\tau),j(\tau))}(\tau) \right],
  \end{equation}
where the forgetting factor $\lambda$ is close to, but less then one, and update $\alpha_i$ at each time step using 
\begin{equation}
\alpha_k(t) = \min\left\{1,\max\{0,\alpha_k(t-1) - \mu(t) \Delta_{B_k}(t)\}\right\} \label{eq:alpha_subgrad_update}
\end{equation}
for some suitable choice of step size $\mu(t)$. Note that $\Delta_{B_k}(t)$ only will be an approximation of the true subgradient, since it is influenced by several earlier values of $\alpha_k(t)$ and also since it is stochastic. Still, numerical experiments have always shown convergence. 

Since keeping all the buffers stable is a necessary optimality condition, an alternative approach is to use the back-pressure algorithm~\cite{Tassiulas:1992,GeorgiadisNT:2006}. Assume a stationary stochastic arrival process with a rate less than or equal to the achievable source-destination rate and let $B_{\mathcal{S}}(t)$ denote the source buffer occupancy at time $t$. Then, a standard derivation of the back-pressure algorithm, using the Lyapunov function $B_{\mathcal{S}}^2(t) + B_i^2(t)$, gives the following design criterion,
\begin{subequations}
  \label{eq:backpressure}
  \begin{IEEEeqnarray}{cl}
    \underset{\{\mathbf{u}_{i},\mathbf{w}_{j}, i, j\}}{\mbox{max}}~ & (B_{\mathcal{S}}(t)-B_i(t)) C_{\mathcal{S}i}^{(i,j)} + B_j(t) C_{j\mathcal{D}}^{(i,j)} \\
    \mbox{s.t. } & \eqref{eq:BF_norms}\quad \&\quad \eqref{eq:constraint_neq}
  \end{IEEEeqnarray}
\end{subequations}
Dividing this criterion by $B_{\mathcal{S}}(t)$ gives an expression of exactly the same form as~\eqref{eq:weighted_opt}, if we set $\alpha_i=1-B_i(t)/B_{\mathcal{S}}(t)$. Therefore, an alternative approach to determine the optimal $\alpha_i$, is to run the back-pressure algorithm until it reaches stationarity, and use a time average of $1-B_i(t)/B_{\mathcal{S}}(t)$ as $\alpha_i$.


Throughout this paper, we consider an exhaustive search under global channel state information (CSI) and buffer state information (BSI) for all proposed schemes to obtain the optimal performance in RS.
A discussion on the complexity of our proposed schemes for both RS and BF aspects is provided in Section~\ref{subsec:complexity}

The conventional centralized/distributed RS approaches \cite{BKR+06JSAC,BSW07TWC,NMV+13CAMAD} can be applied for implementation of the proposed RS schemes.
To reduce the amount of CSI feedback\footnote{Since the proposed schemes require up-to-date CSI, they are primarily applicable in scenarios with low mobility, relative to the symbol duration, such as typical WiFi deployments.}, distributed RS approaches are more desirable than the centralized approach in practice. For example, first of all, the relays can estimate channels based on orthogonal pilot signals from the source and the destination at the same time. Then, the relays can estimate inter-relay channels in a round robin manner using orthogonal MIMO pilot signals.
If BSIs and CSIs for $\SR$ links (i.e., $\mathbf{h}_{\mathcal{S}i}$'s) are shared among the relays, each relay is able to select the local-best receiving relay by regarding itself as the transmitting relay. 
Similarly to the timer-based distributed RS in \cite{BKR+06JSAC,DYH12TVT}, each relay sets a timer based on an inverse of its local-best objective function value and sends a request-to-send message after timer expiration. Then, the destination sends back a clear-to-send message for the earliest access relay.
If the relay receives the message, it broadcasts the RS information to all the nodes. Afterwards, the source and the transmitting relay start to transmit their own packets.
Through this procedure, the best relay pair can be determined in a distributed manner.
The detailed implementation issues are beyond the scope of this work.

The instantaneous rates in \eqref{eq:backpressure} are determined by effective SINR/SNRs at the receiving and transmitting relays, depending on the transmit and receive beamformers. 
Thus, the transmit and receive beamformers have to be determined separately for each candidate pair of relays.
However, finding the optimal BF vectors for every given relay pair is non-convex and therefore we propose an iterative optimal BF-based RS scheme.
Since the iterative solution requires a high computational complexity, 
several low-complexity suboptimal BF-based RS schemes are also proposed in the rest of this section.
Note that the suboptimal schemes have less complexity in the BF design than the optimal scheme but the same RS protocol complexity. (See Section~\ref{subsec:complexity}.)
For simplicity, we omit the time slot index $t$ in the relay indices $i(t)$ and $j(t)$ hereafter.

\subsection{Proposed Optimal Beamforming-based Relay Selection Scheme}
Obviously, if there is no IRI, maximal ratio combining (MRC) at the receiving relay and maximal ratio transmit (MRT) BF at the transmitting relay, named \emph{IRI-free BF}, are optimal. However, this idealized IRI-free BF is not optimal in the presence of IRI.
In this subsection, we propose an iterative optimal BF-based RS scheme to maximize the average end-to-end rate.
Denoting $f(\gamma) = \log_2(1+\gamma)$, 
the optimization problem \eqref{eq:weighted_opt} for given $\{\alpha_i\}$ and for each given relay pair (i.j) is
\begin{align}
\begin{array}{cl}
\underset{\{\mathbf{u}_{i},\mathbf{w}_{j}, i, j\}}{\mbox{max}} & \quad \alpha f\left(\frac{\rho_{\mathcal{S}}|\mathbf{u}_i^H \mathbf{h}_{\mathcal{S}i}|^2}{\mathbf{u}_i^H \left(\rho_{\mathcal{R}}\mathbf{H}_{ji}\mathbf{w}_{j}\mathbf{w}_{j}^H\mathbf{H}_{ji}^H + \mathbf{I}\right) \mathbf{u}_i}\right)\\
&\quad+ (1-\alpha) f\left(\rho_{\mathcal{R}}\mathbf{h}_{j\mathcal{D}}^H\mathbf{w}_{j}\mathbf{w}_{j}^H \mathbf{h}_{j\mathcal{D}}\right)\\
\mbox{s. t.} & \quad \|\mathbf{u}_i\| \leq 1, \|\mathbf{w}_j\| \leq 1.
\end{array}
\end{align}
Since this optimization problem cannot be solved directly due to non-convexity, we propose an alternating optimization which iterates between (i) \emph{for fixed $\mathbf{w}_j$, optimizing $\mathbf{u}_i$}, and (ii) \emph{for fixed $\mathbf{u}_i$, optimizing $\mathbf{w}_j$}.

For given $\mathbf{w}_j$, the optimal receive beamformer $\mathbf{u}_i$ is obviously given by the MMSE solution, 
\begin{align}
\mathbf{u}_i = c_u \left(\rho_{\mathcal{R}}\mathbf{H}_{ji}\mathbf{w}_{j}\mathbf{w}_{j}^H\mathbf{H}_{ji}^H + \mathbf{I}\right)^{-1} \mathbf{h}_{\mathcal{S}i},
\end{align}
where the scaling factor $c_u$ is selected such that $\|\mathbf{u}_i\|=1$.

For given $\mathbf{u}_i$, optimizing the transmit beamformer $\mathbf{w}_j$ 
is a non-convex problem. Denote $\mathbf{g}_{ji} \triangleq \mathbf{H}_{ji}^H \mathbf{u}_{i}$. The Lagrangian of the optimization problem is
\begin{align}
&\mathcal{L}(\mathbf{w}_j, \lambda) = \alpha f\left( \frac{\rho_{\mathcal{S}}|\mathbf{u}_i^H \mathbf{h}_{\mathcal{S}i}|^2}{ \rho_{\mathcal{R}}\mathbf{g}_{ji}^H  \mathbf{w}_j\mathbf{w}_j^H \mathbf{g}_{ji} +  \|\mathbf{u}_i\|^2 } \right)  \nonumber\\
&\quad  + (1-\alpha) f\left(\rho_{\mathcal{R}}\mathbf{h}_{j\mathcal{D}}^H \mathbf{w}_j \mathbf{w}_j^H \mathbf{h}_{j\mathcal{D}} \right) + \lambda \left( 1 -  \mathbf{w}_j^H \mathbf{w}_j \right),
\end{align}
where the gradient with respect to $\mathbf{w}_{j}$ is obtained by
\begin{align}
&\bigtriangledown_{\mathbf{w}_j}\mathcal{L}(\mathbf{w}_{j}, \lambda) \nonumber\\
&= - \frac{\alpha f^{\prime}(\gamma_{\mathcal{S}i})\rho_{\mathcal{S}}|\mathbf{u}_i^H\mathbf{h}_{\mathcal{S}i}|^2}{ \left( \rho_{\mathcal{R}}\mathbf{g}_{ji}^H \mathbf{w}_j\mathbf{w}_j^H \mathbf{g}_{ji} +  \|\mathbf{u}_i\|^2 \right)^2 } 2\rho_{\mathcal{R}}{\mathbf{g}_{ji}}\mathbf{g}_{ji}^H \mathbf{w}_j \nonumber\\
&\quad+ (1-\alpha) f^{\prime}(\gamma_{j\mathcal{D}}) 2\rho_{\mathcal{R}}\mathbf{h}_{j\mathcal{D}}\mathbf{h}_{j\mathcal{D}}^H \mathbf{w}_j - 2\lambda\mathbf{w}_j,
\end{align}
where $\gamma_{\mathcal{S}i} = \frac{\rho_{\mathcal{S}}|\mathbf{u}_i^H \mathbf{h}_{\mathcal{S}i}|^2}{ \rho_{\mathcal{R}}\mathbf{g}_{ji}^H  \mathbf{w}_j\mathbf{w}_j^H \mathbf{g}_{ji}  + \|\mathbf{u}_i\|^2 }$ and $\gamma_{j\mathcal{D}} = \rho_{\mathcal{R}}\mathbf{h}_{j\mathcal{D}}^H\mathbf{w}_j\mathbf{w}_j^H \mathbf{h}_{j\mathcal{D}}$.
Hence, the KKT conditions give
\begin{align}
\left(\lambda\mathbf{I} + \mu {\mathbf{g}_{ji}}\mathbf{g}_{ji}^H\right)\mathbf{w}_j 
= (1-\alpha_j) f^{\prime}(\gamma_{j\mathcal{D}}) \rho_{\mathcal{R}} \mathbf{h}_{j\mathcal{D}} \mathbf{h}_{j\mathcal{D}}^H \mathbf{w}_j,
\label{eq:Lambda}
\end{align}
where $\mu \triangleq \frac{\alpha_i f^{\prime}(\gamma_{\mathcal{S}i})\rho_{\mathcal{S}}|\mathbf{u}_i^H\mathbf{h}_{\mathcal{S}i}|^2}{ \left( \rho_{\mathcal{R}}\mathbf{g}_{ji}^H \mathbf{w}_j{\mathbf{w}_j}^H \mathbf{g}_{ji} +  \|\mathbf{u}_i\|^2 \right)^2 }\rho_{\mathcal{R}}$.
Thus, the optimal $\mathbf{w}_j$ has the form
\begin{align}
\mathbf{w}_j = c_w \left(\lambda\mathbf{I} + \mu {\mathbf{g}_{ji}}\mathbf{g}_{ji}^H\right)^{-1} \mathbf{h}_{j\mathcal{D}},
\label{eq:sol_opt_w}
\end{align}
for some values of the positive real-valued parameters $\lambda$ and $\mu$, and scaling constant $c_w$.
Using the matrix inversion lemma,\footnote{$(A-UD^{-1}V)^{-1} = A^{-1} + A^{-1}U\left(D-VA^{-1}U\right)^{-1}VA^{-1}$ \cite{HJ13MA}.} \eqref{eq:sol_opt_w} is rewritten as
\begin{align}
\mathbf{w}_j &= c_w \left( \lambda^{-1}\mathbf{I} - \lambda^{-1}\mathbf{g}_{ji} \left( \mathbf{g}_{ji}^H \mathbf{g}_{ji} + \lambda/\mu \right)^{-1} \mathbf{g}_{ji}^H\right) \mathbf{h}_{j\mathcal{D}} \nonumber\\
&= \frac{c_w}{\lambda} \left( \mathbf{h}_{j\mathcal{D}} - \frac{\mathbf{g}_{ji}^H \mathbf{h}_{j\mathcal{D}}}{\mathbf{g}_{ji}^H \mathbf{g}_{ji} + \lambda/\mu} \mathbf{g}_{ji}\right) \nonumber\\
&= \frac{c_w}{\lambda} \Bigg( \underbrace{ \mathbf{h}_{j\mathcal{D}} - \frac{\mathbf{g}_{ji}^H \mathbf{h}_{j\mathcal{D}}}{\mathbf{g}_{ji}^H \mathbf{g}_{ji}} \mathbf{g}_{ji} }_{\triangleq  \tilde{\mathbf{w}}^{\perp}_{j}  } \nonumber\\
&\hspace{10mm}+ \left(1 - \frac{\mathbf{g}_{ji}^H \mathbf{g}_{ji}}{\mathbf{g}_{ji}^H \mathbf{g}_{ji} + \lambda/\mu} \right)\underbrace{ \frac{\mathbf{g}_{ji}^H \mathbf{h}_{j\mathcal{D}}}{\mathbf{g}_{ji}^H \mathbf{g}_{ji}} \mathbf{g}_{ji} }_{\triangleq  \tilde{\mathbf{w}}^{\parallel}_{j}  }  \Bigg),
\label{eq:sol_opt_w_2}
\end{align}
where $ \tilde{\mathbf{w}}^{\perp}_{j} $ is defined as the projection of $\mathbf{h}_{j\mathcal{D}}$ onto the orthogonal subspace of $\mathbf{g}_{ji}$ and $ \tilde{\mathbf{w}}^{\parallel}_{j}$ is defined as the projection of $\mathbf{h}_{j\mathcal{D}}$ onto $\mathbf{g}_{ji}$.
Let us define the corresponding normalized vectors ${\mathbf{w}}_j^{\parallel}\triangleq\frac{\tilde{\mathbf{w}}_j^{\parallel}}{\|\tilde{\mathbf{w}}_j^{\parallel}\|}$ and ${\mathbf{w}}_j^{\perp}\triangleq\frac{\tilde{\mathbf{w}}_j^{\perp}}{\|\tilde{\mathbf{w}}_j^{\perp}\|}$.
Since $\|\mathbf{w}_j\|=1$ at the optimum and the scaling constants in \eqref{eq:sol_opt_w_2} are real valued and positive, the optimal  $\mathbf{w}_j$ has the form
\begin{align}
\mathbf{w}_j = \beta {\mathbf{w}}_j^{\parallel} + \sqrt{1-\beta^2}{\mathbf{w}}_j^{\perp},
\label{eq:opt_w}
\end{align}
for some value of $\beta\in[0,1]$. Thus, the optimization problem at hand is reduced to find the optimal $\beta$ parameter for given $(i,j)$ relay pair.
Substituting \eqref{eq:opt_w} into the original objective function gives
\begin{align}
&\underset{\beta\in[0,1]}{\mbox{max}} \quad \alpha f\left(\frac{\rho_{\mathcal{S}}|\mathbf{u}_i^H \mathbf{h}_{\mathcal{S}i}|^2}{\rho_{\mathcal{R}} |\beta \mathbf{g}_{ji}^H {\mathbf{w}}_{j}^{\parallel} |^2 +  \|\mathbf{u}_i\|^2}\right)\nonumber\\
&\hspace{12mm} + (1-\alpha)  f\left(\rho_{\mathcal{R}} |\beta \mathbf{h}_{j\mathcal{D}}^H {\mathbf{w}}_j^{\parallel} + \sqrt{1-\beta^2} \mathbf{h}_{j\mathcal{D}}^H {\mathbf{w}}_j^{\perp} |^2  \right),
\label{eq:obj_function}
\end{align}
where $f(\gamma)=\log_2\left(1+\gamma\right)$.
In general, this objective function may have multiple optima in the interval $[0,1]$ with respect to $\beta$. However, a rough grid search in this interval and a few Gauss-Newton steps starting at the optimum grid point after the grid search can quickly find the global optimum. Note that \eqref{eq:obj_function} is easily vectorized which can speed up the grid search.

Equations \eqref{eq:sol_opt_w_2} and \eqref{eq:obj_function} are iterated to find the optimal beamformers pair until a fixed number of iterations is reached or a certain stopping condition is satisfied.\footnote{In numerical results, we used a stopping condition where the differences in vector norms should be under a certain error tolerance ($\varepsilon_t$), i.e., $\|\mathbf{u}_i(n)-\mathbf{u}_i(n-1)\|<\varepsilon_t$ and $\|\mathbf{w}_j(n)-\mathbf{w}_j(n-1)\|<\varepsilon_t$ for the $n$-th iteration. For all results, we set $\varepsilon_t=10^{-4}$ and then the number of iterations is about several hundreds depending on channel realizations.}

Let the optimized beamformers and $\beta$ parameter be denoted by $\mathbf{u}_i^{\star}$, $\mathbf{w}_j^{\star}$, and $\beta^{\star}$. Then, the instantaneous SNRs for the $\Si$ and $\jD$ links are expressed, respectively, as:
\begin{align}
\gamma_{\mathcal{S}i}^{(i,j)}(t) &= \frac{\rho_{\mathcal{S}}|\mathbf{u}_i^{\star H} \mathbf{h}_{\mathcal{S}i}|^2}{\rho_{\mathcal{R}} |\beta^{\star} \mathbf{g}_{ji}^H {\mathbf{w}}_{j}^{\parallel \star} |^2 +  \|\mathbf{u}_i^{\star}\|^2},\label{eq:sinr_si_optbf}\\
\gamma_{j\mathcal{D}}^{(i,j)}(t) &=\rho_{\mathcal{R}} \left|\beta^{\star} \mathbf{h}_{j\mathcal{D}}^H {\mathbf{w}}_j^{\parallel \star} + \sqrt{1-\beta^{\star 2}} \mathbf{h}_{j\mathcal{D}}^H {\mathbf{w}}_j^{\perp \star} \right|^2,\label{eq:sinr_jd_optbf}
\end{align}
where $\mathbf{w}_j^{\star} = \beta^{\star} {\mathbf{w}}_j^{\parallel \star} + \sqrt{1-\beta^{\star 2}} {\mathbf{w}}_j^{\perp \star}$.
Hence, substituting \eqref{eq:sinr_si_optbf} and \eqref{eq:sinr_jd_optbf} into \eqref{eq:C_si} and \eqref{eq:C_jd}, respectively, the best relay pair is selected by \eqref{eq:backpressure}.

\subsection{Proposed Relay Selection Scheme with Zero-Forcing Beamforming (ZFBF)-based IRI Cancellation}

In this subsection, we propose to optimize a transmit beamformer based on zero-forcing (ZF) at the transmitting relay.
First of all, we use the MRC beamformer for the receiving relay $i$, i.e., $\mathbf{u}_i = \frac{{\mathbf{h}_{\mathcal{S}i}}}{\| \mathbf{h}_{\mathcal{S}i} \|}$ and then maximize the effective channel power gain of the $\RD$ link under a ZF condition.
Therefore, for a given relay pair $(i,j)$, the following optimization problem is formulated:
\begin{subequations}
\begin{IEEEeqnarray}{cl}
\mbox{max} &\quad |\mathbf{w}_{j}^H {\mathbf{h}_{j\mathcal{D}}}|^2 \label{eq:opt_null_space}\\
\mbox{s. t.} &\quad \mathbf{u}_i^H\mathbf{H}_{ji}\mathbf{w}_{j} = 0,\label{eq:opt_null_space_c1}\\
&\quad \|\mathbf{w}_{j}\|=1. \label{eq:opt_null_space_c2}
\end{IEEEeqnarray}
\end{subequations}
Let $\mathbf{V}_{ji}\in\mathbb{C}^{M\times(M-1)}$ be a matrix whose columns span the null-space of $\mathbf{g}_{ji} \triangleq \mathbf{H}_{ji}^H\mathbf{u}_i$. 
Then, any BF vector ${\mathbf{w}_{j}}$ fulfilling the first constraint in \eqref{eq:opt_null_space_c1} can be written as ${\mathbf{w}_{j}}=\mathbf{V}_{ji} \mathbf{z}$, where $\mathbf{z} \in C^{(M-1)\times 1}$.
Hence, the optimization problem is reformulated by
\begin{subequations}
\begin{IEEEeqnarray}{cl}
\underset{\mathbf{z}}{\mbox{max}} & \quad |{\mathbf{z}}^H \mathbf{V}_{ji}^H {\mathbf{h}_{j\mathcal{D}}}|^2 \label{eq:opt_null_space_re}\\
\mbox{s. t.} & \quad \|\mathbf{V}_{ji} \mathbf{z}\|=1,
\end{IEEEeqnarray}
\end{subequations}
which is equivalent to
\begin{subequations}
\begin{IEEEeqnarray}{cl}
\underset{\mathbf{z}}{\mbox{max}} &\quad \displaystyle \frac{\mathbf{z}^H \mathbf{V}_{ji}^H {\mathbf{h}_{j\mathcal{D}}} \mathbf{h}_{j\mathcal{D}}^H {\mathbf{V}_{ji}} \mathbf{z}}{\mathbf{z}^H \mathbf{V}_{ji}^H{\mathbf{V}_{ji}} \mathbf{z}} \label{eq:opt_null_space_final}\\
\mbox{s. t.} &\quad \|\mathbf{V}_{ji} \mathbf{z}\|=1.
\end{IEEEeqnarray}
\end{subequations}
The solution of this problem is $\mathbf{z}^{\star}=c_z (\mathbf{V}_{ji}^H\mathbf{V}_{ji})^{-1}\mathbf{V}_{ji}^H {\mathbf{h}_{j\mathcal{D}}}$, resulting in $\mathbf{w}_j^{\star} = c_z \mathbf{V}_{ji} (\mathbf{V}_{ji}^H \mathbf{V}_{ji})^{-1}$ $\cdot\mathbf{V}_{ji}^H \mathbf{h}_{j\mathcal{D}} =  c_z \tilde{\mathbf{w}}_{j}^{\star} $, where the scalar $c_z$ is chosen so that $\|\mathbf{w}_j^{\star}\|=1$, i.e. $ \mathbf{w}_j^{\star} = \frac{\tilde{\mathbf{w}}_{j}^{\star}}{\|{\tilde{\mathbf{w}}_{j}}^{\star}\|} $ in which
\begin{align}
 \tilde{\mathbf{w}}_{j}^{\star}  &= \mathbf{V}_{ji}(\mathbf{V}_{ji}^H\mathbf{V}_{ji})^{-1}\mathbf{V}_{ji}^H {\mathbf{h}_{j\mathcal{D}}}
 = \left( \mathbf{I} -  \frac{\mathbf{g}_{ji} \mathbf{g}_{ji}^H}{\mathbf{g}_{ji}^H\mathbf{g}_{ji}} \right){\mathbf{h}_{j\mathcal{D}}} \nonumber\\
 &= {\mathbf{h}_{j\mathcal{D}}} - c \mathbf{H}_{ji}^H\mathbf{h}_{\mathcal{S}i},
\label{eq:q_ZFBF}
\end{align}
where the scalar value $c=\frac{ \mathbf{h}_{\mathcal{S}i}^H\mathbf{H}_{ji} {\mathbf{h}_{j\mathcal{D}}}}{ \|\mathbf{h}_{\mathcal{S}i}^H\mathbf{H}_{ji}\|^2 }$ since $\mathbf{u}_{i} = \frac{\mathbf{h}_{\mathcal{S}i}}{\|\mathbf{h}_{\mathcal{S}i}\|}$.
This optimum solution implies a projection of ${\mathbf{h}_{j\mathcal{D}}}$ onto the null-space $\mathbf{V}_{ji}$.

Therefore, substituting $\mathbf{u}_i$ and $\mathbf{w}_j$ into \eqref{eq:SINRi} and \eqref{eq:SNRD}, the instantaneous SNRs for the $\Si$ and $\jD$ links are expressed, respectively, as:
\begin{align}
\gamma_{\mathcal{S}i}^{(i,j)}(t) &= \rho_{\mathcal{S}} \|\mathbf{h}_{\mathcal{S}i}\|^2,\label{eq:sinr_si_zfbf}\\
\gamma_{j\mathcal{D}}^{(i,j)}(t) &=\frac{ \rho_{\mathcal{R}}  \left| \|{\mathbf{h}_{j\mathcal{D}}}\|^2 - \tilde{c} \right|^2}{\|{\mathbf{h}_{j\mathcal{D}}} - c \mathbf{H}_{ji}^H\mathbf{h}_{\mathcal{S}i}\|^2},\label{eq:sinr_jd_zfbf}
\end{align}
where the scalar values $c=\frac{ \mathbf{h}_{\mathcal{S}i}^H\mathbf{H}_{ji} {\mathbf{h}_{j\mathcal{D}}}}{ \|\mathbf{h}_{\mathcal{S}i}^H\mathbf{H}_{ji}\|^2 }$ and $\tilde{c}=\frac{ |\mathbf{h}_{\mathcal{S}i}^H\mathbf{H}_{ji} {\mathbf{h}_{j\mathcal{D}}}|^2 }{ \|\mathbf{h}_{\mathcal{S}i}^H\mathbf{H}_{ji} \|^2 }$.
As a result, substituting \eqref{eq:sinr_si_zfbf} and \eqref{eq:sinr_jd_zfbf} into \eqref{eq:C_si} and \eqref{eq:C_jd}, respectively, the best relay pair is selected by \eqref{eq:backpressure}.

\begin{proposition}
\label{prop:ZFBF_antennas}
The proposed ZFBF-based RS scheme asymptotically achieves the average end-to-end rate of ideal FD relaying with probability one as the number of antennas ($M$) goes to infinity.
\begin{proof}
Let $(\mathbf{H})_m$ denote the $m$-th column vector of the matrix $\mathbf{H}$ and $(\mathbf{H})_{m,l}$ denote the $(m,l)$-th element of the matrix $\mathbf{H}$.
Denoting $\mathbf{g}_{ji} = [g_{ji}^{1},g_{ji}^{2},\ldots,g_{ji}^{M}]^T$ in \eqref{eq:q_ZFBF}, $g_{ji}^{m}=(\mathbf{H}_{ji})_m^H \mathbf{u}_{i}\sim\mathcal{CN}(0,\sigma_{\mathcal{RR}}^2)$ since $(\mathbf{H}_{ji})_{m,l}\sim\mathcal{CN}(0,\sigma_{\mathcal{RR}}^2)$ and $\|\mathbf{u}_i\|=1$. Assuming $\sigma_{\mathcal{RR}}^2=1$ without loss of generality, $g_{ji}^{m}\sim\mathcal{CN}(0,1)$ and therefore $\mathbf{g}_{ji}^H \mathbf{g}_{ji} = \sum_{m=1}^{M} |g_{ji}^{m}|^2$ follows a chi-squared distribution with $2M$ degrees of freedom, i.e., $\mathbf{g}_{ji}^H \mathbf{g}_{ji} \sim \chi_{2M}^{2}$. Meanwhile, the diagonal elements, $(\mathbf{g}_{ji}\mathbf{g}_{ji}^H)_{m,m}\triangleq |g_{ji}^{m}|^2$, follow an exponential distribution with parameter one, i.e., $(\mathbf{g}_{ji}\mathbf{g}_{ji}^H)_{m,m}\sim\mathrm{Exp}(1), \forall m\in\{1,\ldots,M\}$ and the off-diagonal elements, $(\mathbf{g}_{ji}\mathbf{g}_{ji}^H)_{m,l}, \forall m\neq l$, follow the distribution of a product of two independent Gaussians with zero mean and unit variance (see \cite{OM12TSP}). The distribution of each element in $\mathbf{g}_{ji}\mathbf{g}_{ji}^H$ is not varying with respect to $M$ while only the size of matrix grows according to $M$. As a result, as $M\rightarrow\infty$, the denominator $\mathbf{g}_{ji}^H \mathbf{g}_{ji}$ goes to infinity, while all the elements in the numerator $\mathbf{g}_{ji}\mathbf{g}_{ji}^H$ remain as constant values with respect to $M$. Hence, $\frac{\mathbf{g}_{ji}\mathbf{g}_{ji}^H}{\mathbf{g}_{ji}^H \mathbf{g}_{ji}} \rightarrow \mathbf{0}$ as $M\rightarrow \infty$. Accordingly, $\mathbf{w}_{j}^{\star} \rightarrow \frac{\mathbf{h}_{j\mathcal{D}}}{\|\mathbf{h}_{j\mathcal{D}}\|}$ and $\gamma_{j\mathcal{D}}^{(i,j)}(t) \rightarrow \rho_{\mathcal{R}}\|\mathbf{h}_{j\mathcal{D}}\|^2$ as $M \rightarrow \infty$, which completes the proof.
\end{proof}
\end{proposition}


\begin{remark}\label{rm:ZFBF_relays}
The proposed ZFBF-based RS scheme also approach the average end-to-end rate of ideal FD relaying as the number of relays ($K$) goes to infinity due to increased selection diversity. However, increasing the number of relays cannot guarantee to achieve the performance of the ideal FD relaying as in Proposition~\ref{prop:ZFBF_antennas} since its selection diversity is always less than or equal to that of the ideal FD relaying. In other words, the best transmitting relay should meet $\frac{\mathbf{g}_{ji}\mathbf{g}_{ji}^H}{\mathbf{g}_{ji}^H \mathbf{g}_{ji}} = \mathbf{0}$ for achieving the same performance as the ideal FD relaying. Although there exist certain relays satisfying $\frac{\mathbf{g}_{ji}\mathbf{g}_{ji}^H}{\mathbf{g}_{ji}^H \mathbf{g}_{ji}} \approx \mathbf{0}$ with high probability as $K$ goes to infinity, they are a subset of the set of relays while the ideal FD relaying can always take the full selection diversity due to no IRI assumption.
\end{remark}

\begin{remark}\label{rm:optBF}
If the ZFBF solution in \eqref{eq:q_ZFBF} is set to the initial vector for $\mathbf{w}_j$ in the first step of the proposed iterative optimal BF, it always yields a better solution than the ZFBF, regardless of the number of iterations, since $\underset{\mathbf{u}_i}{\mathrm{argmax}}\frac{\rho_{\mathcal{S}}\mathbf{u}_i^H\mathbf{h}_{\mathcal{S}i} \mathbf{h}_{\mathcal{S}i}^H {\mathbf{u}_i}}{\mathbf{u}_i^H \left(\rho_{\mathcal{R}}\mathbf{H}_{ji}\mathbf{w}_{j}\mathbf{w}_{j}^H\mathbf{H}_{ji}^H + \mathbf{I}\right) {\mathbf{u}_i}}=\underset{\mathbf{u}_i}{\mathrm{argmax}}\frac{\rho_{\mathcal{S}}\mathbf{u}_i^H\mathbf{h}_{\mathcal{S}i} \mathbf{h}_{\mathcal{S}i}^H {\mathbf{u}_i}}{\mathbf{u}_i^H{\mathbf{u}_i}}=\frac{\mathbf{h}_{\mathcal{S}i}}{\|\mathbf{h}_{\mathcal{S}i}\|}$ which yields exactly the same BF pair of the ZFBF-based RS scheme and an additional iteration gives a better solution. 
We therefore propose to initialize the alternating optimization in this way, even though there are no guarantees that this will provide the global optimum.
\end{remark}


\subsection{Proposed Relay Selection Scheme with Minimum Mean Square Error (MMSE)-based IRI Suppression}
In this subsection, we propose a receive beamformer based on minimum mean square error (MMSE) for maximizing the effective SINR at the receiving relay.
First of all, we use the MRT beamformer for the transmitting relay $j$, i.e., $\mathbf{w}_j = \frac{{\mathbf{h}_{j\mathcal{D}}}}{\| \mathbf{h}_{j\mathcal{D}} \|}$, and then we find a receive beamformer for maximizing the effective SINR at the receiving relay.
Therefore, for given relay pair $(i,j)$ and $\mathbf{w}_j$, the following optimization problem is formulated:
\begin{subequations}
\label{eq:OPT_maxSNR}
\begin{IEEEeqnarray}{cl}
\underset{\mathbf{u}_i}{\mbox{max}} &\quad \frac{{ \rho_{\mathcal{S}} }\mathbf{u}_i^H \mathbf{h}_{\mathcal{S}i} \mathbf{h}_{\mathcal{S}i}^H {\mathbf{u}_i}}{\mathbf{u}_i^H \left({ \rho_{\mathcal{R}} } \mathbf{H}_{ji}\mathbf{w}_{j}\mathbf{w}_{j}^H\mathbf{H}_{ji}^H + { \mathbf{I} }\right) {\mathbf{u}_i}}\\
\mbox{s. t.} & \quad \|\mathbf{u}_{i}\|=1.
\end{IEEEeqnarray}
\end{subequations}
The solution of \eqref{eq:OPT_maxSNR} is given by the scaled MMSE as $\mathbf{u}_i = c_{m}\left(\rho_{\mathcal{R}}\mathbf{H}_{ji}\mathbf{w}_{j}\mathbf{w}_{j}^H\mathbf{H}_{ji}^H + \mathbf{I}\right)^{-1}\mathbf{h}_{\mathcal{S}i}$ where the scaling factor $c_{m}$ is chosen such that $\|\mathbf{u}_{i}\|=1$. 

Therefore, substituting $\mathbf{u}_i$ and $\mathbf{w}_j$ into \eqref{eq:SINRi} and \eqref{eq:SNRD}, the instantaneous SNRs for the $\Si$ and $\jD$ links are expressed, respectively, as:
\begin{align}
\gamma_{\mathcal{S}i}^{(i,j)}(t) 
&= {\rho_{\mathcal{S}} \mathbf{h}_{\mathcal{S}i}^H\left(\frac{\rho_{\mathcal{R}}}{\|\mathbf{h}_{j\mathcal{D}}\|^2}\mathbf{H}_{ji}\mathbf{h}_{j\mathcal{D}}\mathbf{h}_{j\mathcal{D}}^H\mathbf{H}_{ji}^H + \mathbf{I}\right)^{-1}\mathbf{h}_{\mathcal{S}i},} \label{eq:sinr_si_mmse}\\
\gamma_{j\mathcal{D}}^{(i,j)}(t) &=\rho_{\mathcal{R}} \|\mathbf{h}_{j\mathcal{D}}\|^2.\label{eq:sinr_jd_mmse}
\end{align}
Substituting \eqref{eq:sinr_si_mmse} and \eqref{eq:sinr_jd_mmse} into \eqref{eq:C_si} and \eqref{eq:C_jd}, respectively, the best relay pair is selected by \eqref{eq:backpressure}.


\begin{proposition}
\label{prop:MMSE_lowSNR}
The proposed MMSE-based RS scheme asymptotically achieves the average end-to-end rate of ideal FD relaying at low SNR.
\begin{proof}
As $\rho_R \to 0$, $\gamma_{\mathcal{S}i}^{(i,j)}(t) \to \rho_{\mathcal{S}}\mathbf{h}_{\mathcal{S}i}^H \mathbf{I}^{-1}\mathbf{h}_{\mathcal{S}i}=\rho_{\mathcal{S}}\|\mathbf{h}_{\mathcal{S}i}\|^2$, which completes the proof.
\end{proof}
\end{proposition}

\subsection{Proposed Relay Selection Scheme with Orthonormal Basis (OB)-based IRI Cancellation}

In this subsection, we propose a perfect IRI cancellation BF based on orthonormal basis vectors.
To this end, we first generate two random orthonormal vectors $\mathbf{u}$ and $\mathbf{q}$, i.e., $\mathbf{u}^H\mathbf{q}=0$, $\Vert\mathbf{u}\Vert=1$, and $\Vert\mathbf{q}\Vert=1$.
Then, we use $\mathbf{u}$ as the receive beamformer at the receiving relay and $\mathbf{w}_{j}=\frac{\mathbf{H}_{ji}^{-1}\mathbf{q}}{\Vert\mathbf{H}_{ji}^{-1}\mathbf{q}\Vert}$ as the transmit beamformer at the transmitting relay, respectively.


Since $\mathbf{u}^H \mathbf{H}_{ji} \mathbf{w}_{j}=\mathbf{u}^H \mathbf{H}_{ji} \frac{\mathbf{H}_{ji}^{-1} \mathbf{q}}{\Vert\mathbf{H}_{ji}^{-1} \mathbf{q}\Vert}=0$, $\|\mathbf{u}\|=1$, and $\|\mathbf{w}_j\|=1$, substituting $\mathbf{u}$ and $\mathbf{w}_j$ into \eqref{eq:SINRi} and \eqref{eq:SNRD}, the instantaneous SNRs for the $\Si$ and $\jD$ links are expressed as:
\begin{align}
\gamma_{\mathcal{S}i}^{(i,j)}(t) &= \rho_{\mathcal{S}} |\tilde{h}_{\mathcal{S}i}|^2,\label{eq:sinr_si_ob}\\
\gamma_{j\mathcal{D}}^{(i,j)}(t) &= \rho_{\mathcal{R}}|\tilde{h}_{j\mathcal{D}}|^2,\label{eq:sinr_jd_ob}
\end{align}
where $\tilde{h}_{\mathcal{S}i} = \mathbf{u}^H \mathbf{h}_{\mathcal{S}i} \sim \mathcal{CN}(0,\sigma_{\mathcal{S}\mathcal{R}}^2)$ and $\tilde{h}_{j\mathcal{D}} = \mathbf{h}_{j\mathcal{D}}^H\mathbf{w}_{j} \sim \mathcal{CN}(0,\sigma_{\mathcal{R}\mathcal{D}}^2)$.
As a result, substituting \eqref{eq:sinr_si_ob} and \eqref{eq:sinr_jd_ob} into \eqref{eq:C_si} and \eqref{eq:C_jd}, respectively, the best relay pair is selected by \eqref{eq:backpressure}.

\begin{remark}
From \eqref{eq:sinr_si_ob} and \eqref{eq:sinr_jd_ob}, the proposed OB-based RS scheme achieves the average end-to-end rate of ideal FD relaying with a single antenna at the relays. 
\end{remark}

\begin{remark}\label{rm:ZFBF_OB}
The proposed ZFBF-based RS scheme is always better than the proposed OB-based RS scheme in terms of the average end-to-end rate while both of the schemes perfectly cancel IRI. However, it is not always better in the viewpoint of instantaneous RS and BF, since the randomly chosen beamformers in the proposed OB-based RS scheme might be better at a certain RS instance for not the $\SR$ link rate but the weighted sum-rate.
\end{remark}

\subsection{Proposed SINR-based Relay Selection Scheme with Beamforming Neglecting IRI}

Although the IRI-free BF is not optimal in the presence of IRI, we propose to use them and utilize effective SINR/SNR measures after BF in RS as the simplest joint RS and BF scheme.
Accordingly, for relay pair $(i,j)$, the receive BF vector is given by $\mathbf{u}_i = \frac{{\mathbf{h}_{\mathcal{S}i}}}{\| \mathbf{h}_{\mathcal{S}i} \|}$
and the transmit BF vector is given by $\mathbf{w}_j = \frac{{\mathbf{h}_{j\mathcal{D}}}}{\Vert \mathbf{h}_{j\mathcal{D}} \Vert}$.
Substituting $\mathbf{u}_i$ and $\mathbf{w}_j$ into \eqref{eq:SINRi} and \eqref{eq:SNRD}, 
the instantaneous SINR and SNR of both the $\Si$ and $\jD$ links are obtained, respectively, by
\begin{align}
\gamma_{\mathcal{S}i}^{(i,j)}(t) 
&=\frac{ \Vert\mathbf{h}_{\mathcal{S}i}\Vert^2 \rho_{\mathcal{S}}}{1 + \displaystyle\frac{|\mathbf{h}_{\mathcal{S}i}^H\mathbf{H}_{ji} {\mathbf{h}_{j\mathcal{D}}} |^2}{\Vert\mathbf{h}_{\mathcal{S}i}\Vert^2\Vert\mathbf{h}_{j\mathcal{D}}\Vert^2}\rho_{\mathcal{R}}  },
\label{eq:sinr_cap_si}\\
\gamma_{j\mathcal{D}}^{(i,j)}(t) &= \rho_{\mathcal{R}} \Vert\mathbf{h}_{j\mathcal{D}}\Vert^2.
\label{eq:sinr_cap_jd}
\end{align} 
Substituting \eqref{eq:sinr_cap_si} and \eqref{eq:sinr_cap_jd} into \eqref{eq:C_si} and \eqref{eq:C_jd}, respectively, the best relay pair is selected by \eqref{eq:backpressure}.

\subsection{Discussion on Complexity of the Proposed Joint Relay Selection and Beamforming}\label{subsec:complexity}

For the RS protocol, an exhaustive search within $K\times(K-1)$ combinations with global CSI and BSI is required to obtain the optimal performance. Accordingly, the complexity of the optimal relay pair selection is $\mathcal{O}(K^2)$ for all the schemes. However, through the distributed RS approach, CSIs for $\RD$ and $\RR$ links can be directly estimated at each relay from pilot signals and thus the amount of feedback on CSI can be reduced from $(2KM + M^2K(K-1)/2)$ to $KM$. It is also worth mentioning that the number of available (fixed) relays in practical network scenarios would not be so large due to geographical limitations. On the other hand, the proposed schemes can still be effective even for a two-relay network if sufficient number of antennas are available at the relays. Moreover, as will be shown in Section \ref{subsec:rate}, the performance enhancement from more than five relays would be marginal.

\Sumin{For BF design for given relay pair, if we denote the complexity for computing a BF vector including $M$ dimensional matrices and/or vectors by $C_{\mathrm{BF}}$, 
the complexity of all the suboptimal BF schemes becomes $2C_{\mathrm{BF}}$ 
since they require to compute two BF vectors.
However, they can have different computational times in practice according to degree of matrix and/or vector computations. For instance, the MMSE-based scheme requires to compute $M$ dimensional matrix inversion, while the SINR-based scheme only requires vector normalizations.
On the other hand, the complexity of the optimal BF scheme is derived as $L(2C_{\mathrm{BF}} + C_{\beta})$
, where $L$ denotes the number of iterations and $C_{\beta}$ denotes the complexity to find the optimal $\beta$ including both a rough line search and some Gauss-Newton steps.}
Consequently, the optimal BF scheme requires additional complexity for iteration process and an additional optimization parameter, compared to the suboptimal BF schemes.
After all, the suboptimal BF schemes are useful for online operation and a network which consists of nodes with low computation power.

\section{Performance Evaluation} \label{sec:performance_evaluation}

In this section, we evaluate the proposed joint buffer-aided RS and BF schemes in terms of the average end-to-end rate and average delay through Monte-Carlo simulations, compared to conventional HD RS schemes and SFD-MMRS scheme representing state-of-the-art in the literature. For multiple-antenna extension of the conventional schemes, we suppose that they use the IRI-free beamformers. 
As an upper bound of the average end-to-end rate, we consider the optimal joint RS and BF in \eqref{eq:weighted_opt} assuming no IRI.
The optimal weight factors $\{\alpha_k^{\star}\}$ are adaptively obtained based on the back-pressure algorithm during a pre-training phase and applied to each relay pair selection.
We assume half-full initial buffer state at all relays for the pre-training phase but zero initial buffer state for data transmission phase.
We consider Rayleigh block fading channels with average gains of $\sigma_{\mathcal{S}\mathcal{R}_i}^2$, $\sigma_{\mathcal{R}_j\mathcal{D}}^2$, and $\sigma_{\mathcal{R}_j\mathcal{R}_i}^2$ for $i,j\in\mathcal{K}$, 10000 packet transmissions from the source, and assume $P_{\mathcal{S}} = P_{\mathcal{R}}$ throughout all simulations. 

\subsection{Benchmarks}
\subsubsection{HD Best Relay Selection (BRS) Scheme \cite{BKR+06JSAC}}
In the HD-BRS scheme without buffering at relays, the best relay is determined by
\begin{align}
i^{\star} &= \underset{i \in \mathcal{K}}{\mathrm{argmax}}~\min\left\{C_{\mathcal{S}i}(t), C_{i\mathcal{D}}(t)\right\},
\end{align}
where $C_{\mathcal{S}i}(t) = \frac{1}{2}\log_2\left(1 + \gamma_{\mathcal{S}i}(t)\right)$ and $C_{i\mathcal{D}}(t) = \frac{1}{2}\log_2\left(1 + \gamma_{i\mathcal{D}}(t)\right)$.

\subsubsection{HD $\mathrm{max-max}$ Relay Selection (MMRS) Scheme \cite{IMS12TWC}}
In the HD-MMRS scheme maintaining the two-phase operation, the best relay at first time slot, $i^{\star}$, and the best relay at second time slot, $j^{\star}$, are selected as follows:
\begin{align}
i^{\star} &= \underset{i\in \mathcal{K}}{\mathrm{argmax}}~C_{\mathcal{S}i}(2t),\quad
j^{\star} = \underset{j\in \mathcal{K}}{\mathrm{argmax}}~C_{j\mathcal{D}}(2t+1),
\end{align}
where $C_{\mathcal{S}i}(2t) = \min \big\{ \frac{1}{2}\log_2\left(1 + \gamma_{\mathcal{S}i}(2t)\right), B_{max} - B_i(2t-1) \big\}$ and $C_{j\mathcal{D}}(2t+1) = \min \big\{ \frac{1}{2}\log_2(1 + \gamma_{j\mathcal{D}}(2t+1)),$ $B_j(2t) \big\}$.

\subsubsection{HD $\mathrm{max-link}$ Relay Selection (MLRS) Scheme \cite{KCT12TWC}}
Since the HD-MLRS scheme has been developed for fixed rate transmission, we slightly modify it to adaptive rate transmission by adding a link selection parameter $d_i$ for the $i$-th relay.
In the conventional HD-MLRS scheme releasing the two-phase operation condition, the best relay $i^{\star}$ and the best link $d_{i^{\star}}$ (integer variable; receiving 1 or transmitting 0) at each time slot are determined by
\begin{align}
\left(i^{\star}, d_{i^{\star}}\right) = \underset{i\in \mathcal{K},d_i\in\{0,1\}}{\mathrm{argmax}} \left\{ d_i  C_{\mathcal{S}i}(t) + (1-d_i) C_{i\mathcal{D}}(t) \right\},
\end{align}
where $C_{\mathcal{S}i}(t) = \min \left\{ \frac{1}{2}\log_2\left(1 + \gamma_{\mathcal{S}i}(t)\right), B_{max} - B_i(t-1) \right\}$ and $C_{i\mathcal{D}}(t) = \min \big\{ \frac{1}{2}\log_2\left(1 + \gamma_{i\mathcal{D}}(t)\right),$ $B_i(t-1) \big\}$.


\subsubsection{SFD-MMRS Scheme \cite{IKS12TVT}}
In the SFD-MMRS scheme, each $\SR$ or  $\RD$ link selects the best relay and the second best relay based on channel gains without consideration of IRI. Denote the relay indices of the best and second best relays by $i_1$ and $j_2$ for the receiving relay and $i_1$ and $j_2$ for the transmitting relay, respectively. Then they are selected as follows: 
\begin{align}
i_1 &= \underset{i\in\mathcal{K}}{\mathrm{argmax}}~C_{\mathcal{S}i}(t), 
\quad i_2 = \underset{i\in\mathcal{K}\backslash\{i_1\}}{\mathrm{argmax}}~C_{\mathcal{S}i}(t), \nonumber\\
j_1 &= \underset{j\in\mathcal{K}}{\mathrm{argmax}}~C_{j\mathcal{D}}(t), 
\quad j_2 = \underset{j\in\mathcal{K}\backslash\{j_1\}}{\mathrm{argmax}}~C_{j\mathcal{D}}(t), \nonumber
\end{align}
where $C_{\mathcal{S}i}(t) = \min \left\{ \log_2\left(1 + \gamma_{\mathcal{S}i}(t)\right), B_{max} - B_i(t-1) \right\}$ and $C_{j\mathcal{D}}(t) = \min \big\{ \log_2\left(1 + \gamma_{j\mathcal{D}}(t)\right),$ $B_j(t-1) \big\}$.
If the best relays for both links are same, it finds the best relay pair among combinations with the second best relays based on a minimum of achievable rates, i.e., 
\begin{align}
(i^{\star}, j^{\star}) = \left\{
\begin{array}{lll}
(i_1, j_1), & \mbox{\text{if $i_1\neq j_1$}}\\
(i_2, j_1), & \mbox{\text{if $i_1= j_1$ and $\min\{C_{\mathcal{S}i_2}(t), C_{j_1\mathcal{D}}(t)\}$}}\\ 
& \hspace{18mm}\mbox{\text{$> \min\{C_{\mathcal{S}i_1}(t), C_{j_2\mathcal{D}}(t)\}$}}\\
(i_1, j_2), & \mbox{\text{otherwise.}}
\end{array}
\right.
\end{align}
Since the SFD-MMRS scheme assumes a single antenna at the relays and no IRI, we extend it to multiple antennas using the IRI-free BF, but include the actual IRI in the results marked by ``SFD-MMRS-IRI''.
The performance degradation of the non-ideal SFD-MMRS scheme due to IRI is shown as numerical results in the following subsections.

\subsection{Average End-to-End Rate}\label{subsec:rate}
\subsubsection{i.i.d. Channel Case ($\sigma_{\mathcal{S}\mathcal{R}_i}^2=\sigma_{\mathcal{R}_j\mathcal{D}}^2=\sigma_{\mathcal{R}_j\mathcal{R}_i}^2=0~\mathrm{dB},\forall i,j\in\mathcal{K}$)}

\begin{figure}[tp]
\centering
     \includegraphics[width=8.6cm]{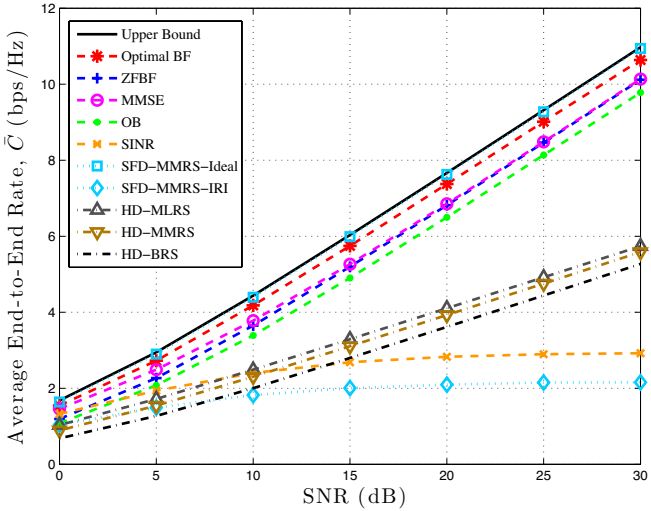}
\vspace{-3mm}     
    \caption{Average end-to-end rate vs. SNR for i.i.d. channel case ($K=2$, $M=2$, $B_{max}\rightarrow\infty$, $\sigma_{\mathcal{S}\mathcal{R}_i}^2=\sigma_{\mathcal{R}_j\mathcal{D}}^2=\sigma_{\mathcal{R}_j\mathcal{R}_i}^2=0~\mathrm{dB},\forall i,j\in\mathcal{K}$)}
\label{fig:C_K2_M2}
\vspace{-5mm}
\end{figure}

Fig.~\ref{fig:C_K2_M2} shows the average end-to-end rate for varying SNR when $K=2$, $M=2$, $B_{max}\rightarrow\infty$, and the average channel qualities of all the links are identical.
The ideal SFD-MMRS scheme performs close to the upper bound obtained by \eqref{eq:weighted_opt}.
This validates that our weighted sum-rate maximization based on instantaneous rates works well to maximize the average end-to-end rate.
If we impose IRI into the SFD-MMRS scheme, its performance is significantly degraded with increasing SNR, i.e., in the interference-limited regime.
Although the proposed SINR-based RS scheme improves the average end-to-end rate, its contribution is not significant at medium/high SNR.
On the contrary, the average end-to-end rates of the other proposed schemes still increase with increasing SNR due to IRI cancellation/suppression.
Since the proposed OB-based RS scheme achieves the single antenna upper bound, the proposed optimal BF-based, ZFBF-based, and MMSE-based RS schemes outperform the single antenna upper bound.
While the optimal BF optimizes both transmit and receive beamformers iteratively, the ZFBF and MMSE BF optimize just one of the beamformers fixing the other beamformer.
As a result, the ZFBF-based and MMSE-based RS schemes achieve lower average end-to-end rates than the optimal BF-based RS scheme. Moreover, the MMSE-based RS scheme achieves better performance at low/medium SNR than the ZFBF-based RS scheme, since the MMSE BF can achieve the ideal upper bound at low SNR regime as proved in Proposition~\ref{prop:MMSE_lowSNR}.

Regarding the conventional HD RS schemes, the HD-MLRS and HD-MMRS schemes outperform the HD-BRS scheme since they additionally utilize buffering at relays. Furthermore, the HD-MLRS scheme outperforms the HD-MMRS scheme since it obtains more diversity gain by releasing the two-phase operation condition.
Compared to the HD RS schemes, the slopes of curves of the proposed optimal BF-based, ZFBF-based, and MMSE-based RS schemes are almost double. For instance, at $\mathrm{SNR}=30$ dB, the proposed optimal BF-based RS scheme excesses twice the average end-to-end rate by the HD-BRS scheme and the proposed ZFBF-based and MMSE-based RS schemes achieve slightly less performance than it.

\begin{figure}[tp]
\centering
\includegraphics[width=8.6cm]{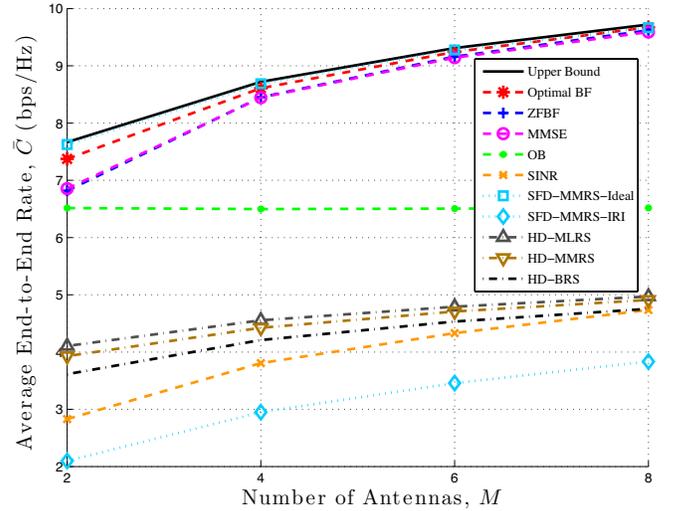}
\vspace{-3mm}
\caption{Effect of the number of antennas ($K=2$, $\mathrm{SNR}=20$ dB, $B_{max}\rightarrow\infty$, $\sigma_{\mathcal{S}\mathcal{R}_i}^2=\sigma_{\mathcal{R}_j\mathcal{D}}^2=\sigma_{\mathcal{R}_j\mathcal{R}_i}^2=0~\mathrm{dB},\forall i,j\in\mathcal{K}$)}
\label{fig:C_varying_M}
\vspace{-5mm}
\end{figure}

Fig.~\ref{fig:C_varying_M} shows the average end-to-end rates for varying number of antennas at the relays when $K=2$, $\mathrm{SNR}=20$ dB, $B_{max}\rightarrow\infty$, and i.i.d. IRI channel conditions.
Except for the proposed OB-based RS scheme, 
the other schemes basically increase the average end-to-end rate as the number of antennas increases.
The proposed optimal BF-based, ZFBF-based, and MMSE-based RS schemes converge to the upper bound when increasing the number of antennas, as shown in Proposition~\ref{prop:ZFBF_antennas} for the ZFBF-based RS scheme and stated in Remark~\ref{rm:optBF} for the optimal BF-based RS scheme.
Eventually, it is worth noting that the proposed optimal BF-based, ZFBF-based, and MMSE-based RS schemes can fully recover the loss of multiplexing gain caused by the HD relaying as the number of antennas increases.
Even if the number of antennas increases, the average end-to-end rates achieved by the SINR-based RS scheme and the non-ideal SFD-MMRS scheme are still less than those by the HD RS schemes due to the effects of strong IRI and limited selection diversity.

\begin{figure}[tp]
\centering
\includegraphics[width=8.6cm]{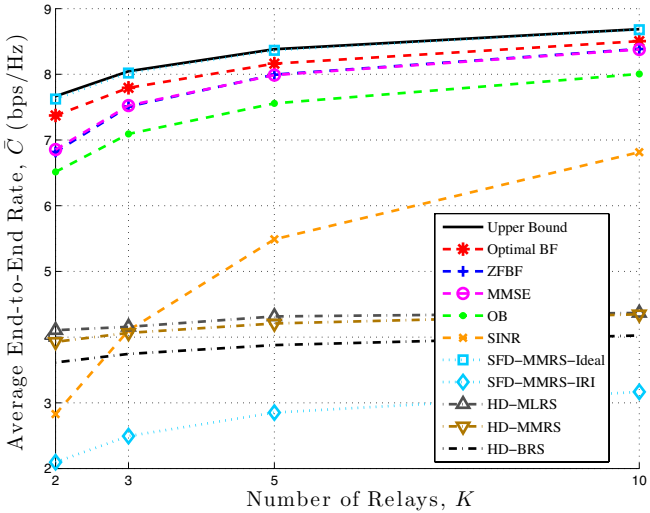}
\vspace{-3mm}
\caption{Effect of the number of relays ($M=2$, $\mathrm{SNR}=20$ dB, $B_{max}\rightarrow\infty$, $\sigma_{\mathcal{S}\mathcal{R}_i}^2=\sigma_{\mathcal{R}_j\mathcal{D}}^2=\sigma_{\mathcal{R}_j\mathcal{R}_i}^2=0~\mathrm{dB},\forall i,j\in\mathcal{K}$)}
\label{fig:C_varying_K}
\vspace{-5mm}
\end{figure}

Fig.~\ref{fig:C_varying_K} shows the average end-to-end rate for varying the number of relays.
All the schemes achieve improved average end-to-end rates due to the increased selection diversity gain as the number of relays increases.
Differently from in Fig.~\ref{fig:C_varying_M}, the proposed SINR-based RS scheme obtains significant gains with increasing number of relays.
Accordingly, it outperforms the conventional HD RS schemes when $K>3$.
In contrast, the non-ideal SFD-MMRS scheme is still the worst with marginal improvements.
This is because it never considers IRI at all while the SINR-based RS scheme takes IRI into account at the RS.
The optimal BF-based, ZFBF-based, and MMSE-based RS schemes also approach the upper bound as the number of relays increases.
Thus, they can asymptotically recover the loss of multiplexing gain with respect to the number of relays even if the exact convergence is not guaranteed in this case.
However, it is shown that the rate improvement is much slower than when increasing the number of antennas. 
When $K=10$, all the proposed schemes except for the SINR-based RS scheme achieve greater than or equal to double the average end-to-end rate of the HD-BRS scheme.

\subsubsection{i.i.d. Channel Cases with Different IRI Intensities} 

To investigate the effect of average IRI intensity, we consider two different average IRI channel conditions: (i) \emph{weak IRI case} ($\sigma_{\mathcal{R}_j\mathcal{R}_i}^2=-10~\mathrm{dB}, \sigma_{\mathcal{S}\mathcal{R}_i}^2=\sigma_{\mathcal{R}_j\mathcal{D}}^2=0~\mathrm{dB}, \forall i,j\in\mathcal{K}$); (ii) \emph{strong IRI case} ($\sigma_{\mathcal{R}\mathcal{R}}^2=10~\mathrm{dB}, \sigma_{\mathcal{S}\mathcal{R}_i}^2=\sigma_{\mathcal{R}_j\mathcal{D}}^2=0~\mathrm{dB}, \forall i,j\in\mathcal{K}$).

\begin{figure}[tp]
\centering
\subfigure[]{
	\includegraphics[width=8.6cm]{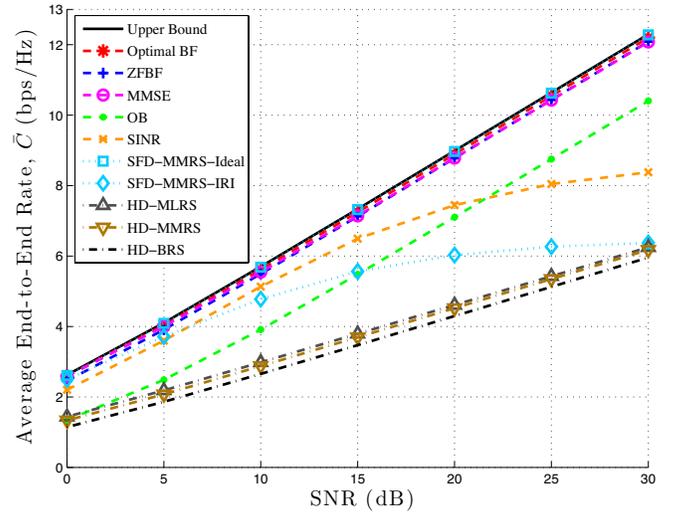}}
    \subfigure[]{
    \includegraphics[width=8.6cm]{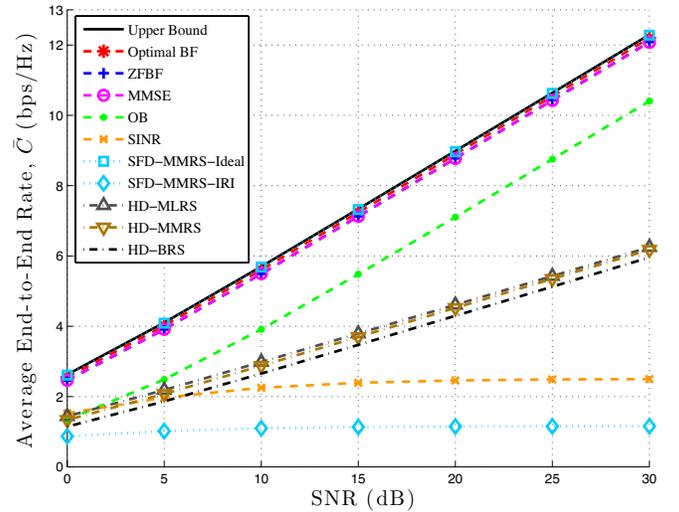}}
\caption{Effect of average IRI intensity ($K=3$, $M=4$, $B_{max}\rightarrow\infty$, $\sigma_{\mathcal{S}\mathcal{R}_i}^2=\sigma_{\mathcal{R}_j\mathcal{D}}^2=0~\mathrm{dB},\forall i,j\in\mathcal{K}$) (a) Weak IRI ($\sigma_{\mathcal{R}_j\mathcal{R}_i}^2=-10~\mathrm{dB},\forall i,j\in\mathcal{K}$) (b) Strong IRI ($\sigma_{\mathcal{R}_j\mathcal{R}_i}^2=10~\mathrm{dB},\forall i,j\in\mathcal{K}$)}
\label{fig:nid_C_K3_M4}
\vspace{-5mm}
\end{figure}

Fig.~\ref{fig:nid_C_K3_M4}~(a) shows the average end-to-end rate for weak IRI case when $K=3$, $M=4$, and $B_{max}\rightarrow\infty$.
When $K=3$ and $M=4$, the proposed optimal BF-based, ZFBF-based, and MMSE-based RS schemes already almost approach the upper bound regardless of SNR.
The optimal BF-based RS scheme outperforms the other schemes and the MMSE-based RS scheme slightly outperforms the ZFBF-based RS scheme at low SNR.
While the OB-based RS scheme still achieves the single antenna upper bound regardless of the average IRI intensity, the SINR-based RS scheme and the non-ideal SFD-MMRS scheme always outperform the HD RS schemes and outperform the OB-based RS scheme at low/medium SNR. Especially, when $\mathrm{SNR}=0$ dB, the non-ideal SFD-MMRS scheme achieves almost the same performance as the ideal upper bound since this channel condition yields very weak interference which is negligible.

Fig.~\ref{fig:nid_C_K3_M4}~(b) shows the average end-to-end rate for strong IRI case in the same setup.
The proposed ZFBF-based and OB-based RS schemes yield exactly the same performance as in Fig.~\ref{fig:nid_C_K3_M4}~(a) since they do not depend on the intensity of IRI due to perfect IRI cancellation. The proposed MMSE-based RS scheme achieves almost identical performance as the proposed ZFBF-based RS scheme even at low SNR since the intensity of IRI is already strong compared to the $\SR$ channel conditions. The optimal BF-based RS scheme performs between the upper bound and the MMSE-based and ZFBF-based RS schemes regardless of the intensity of IRI and SNR. In contrast, the non-ideal SFD-MMRS scheme significantly degrades the average end-to-end rate which is always worse than those of the HD RS schemes. 

\subsubsection{non-i.i.d. Channel Case} 
\begin{figure}[tp]
\centering
\includegraphics[width=8.6cm]{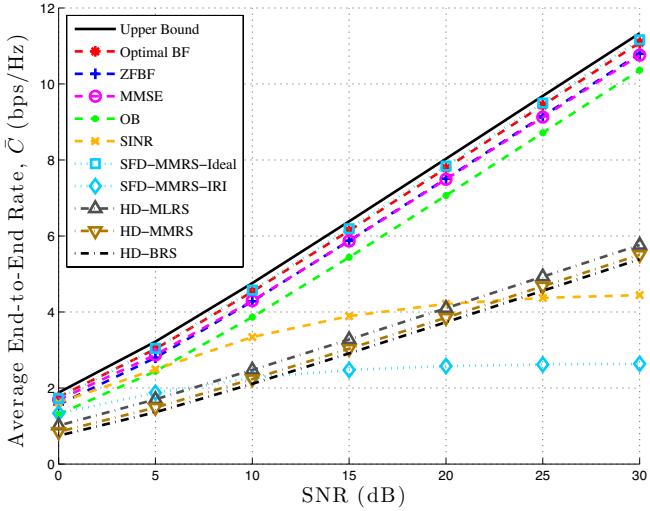}
\vspace{-3mm}
\caption{Average end-to-end rate vs. SNR for non-i.i.d. channel case ($K=3$, $M=2$, $[\sigma_{\mathcal{S}\mathcal{R}_1}^2,\sigma_{\mathcal{S}\mathcal{R}_2}^2,\sigma_{\mathcal{S}\mathcal{R}_3}^2]=[1,0,-1]~\mathrm{dB}$, $[\sigma_{\mathcal{R}_1\mathcal{D}}^2,\sigma_{\mathcal{R}_2\mathcal{D}}^2,\sigma_{\mathcal{R}_3\mathcal{D}}^2]=[-1,0,1]~\mathrm{dB}$, $\sigma_{\mathcal{R}_1\mathcal{R}_2}^2=\sigma_{\mathcal{R}_2\mathcal{R}_1}^2=\sigma_{\mathcal{R}_2\mathcal{R}_3}^2=\sigma_{\mathcal{R}_3\mathcal{R}_2}^2=0~\mathrm{dB},\sigma_{\mathcal{R}_1\mathcal{R}_3}^2=\sigma_{\mathcal{R}_3\mathcal{R}_1}^2=-1~\mathrm{dB}$)}
\label{fig:C_nid}
\vspace{-5mm}
\end{figure}

As a more practical network scenario, we consider a network with three relays where the links have different average channel gains as $[\sigma_{\mathcal{S}\mathcal{R}_1}^2,\sigma_{\mathcal{S}\mathcal{R}_2}^2,\sigma_{\mathcal{S}\mathcal{R}_3}^2]=[1,0,-1]~\mathrm{dB}$, $[\sigma_{\mathcal{R}_1\mathcal{D}}^2,\sigma_{\mathcal{R}_2\mathcal{D}}^2,\sigma_{\mathcal{R}_3\mathcal{D}}^2]=[-1,0,1]~\mathrm{dB}$, and $\sigma_{\mathcal{R}_1\mathcal{R}_2}^2=\sigma_{\mathcal{R}_2\mathcal{R}_1}^2=\sigma_{\mathcal{R}_2\mathcal{R}_3}^2=\sigma_{\mathcal{R}_3\mathcal{R}_2}^2=0~\mathrm{dB},\sigma_{\mathcal{R}_1\mathcal{R}_3}^2=\sigma_{\mathcal{R}_3\mathcal{R}_1}^2=-1~\mathrm{dB}$.
In this setup, $\{\mathcal{S}-\mathcal{R}_2-\mathcal{D}\}$ are equally spaced, $\mathcal{R}_1$ is closer to $\mathcal{S}$ and $\mathcal{R}_3$ is closer to $\mathcal{D}$.
$\{\mathcal{R}_1-\mathcal{R}_2-\mathcal{R}_3\}$ are equally spaced and therefore, $\{\mathcal{R}_1-\mathcal{R}_3\}$ becomes a less interfered channel than $\{\mathcal{R}_1-\mathcal{R}_2\}$ and $\{\mathcal{R}_2-\mathcal{R}_3\}$.
Fig.~\ref{fig:C_nid} shows the average end-to-end rate with various SNR values for the non-i.i.d. channel case.
The basic trend is the same as that for i.i.d. channel cases shown previously. Hence, our proposed weighted sum-rate maximization approach works well also under non-identical channel condition.
Furthermore, the proposed optimal BF-based, ZFBF-based, MMSE-based RS schemes are still effective for generalized asymmetric network topologies.

\subsection{Delay Performance}\label{subsec:delay}

Basically, the buffer-aided relaying obtains additional selection diversity gain by sacrificing delay performance. Even if we mainly focus on the average end-to-end rate of delay-tolerant applications, we evaluate the average delay performance through simulations.
Due to a full-queue assumption at the source, the average delay is defined as the average queueing delay at relays, which implies the time difference between the arrival time of a single packet at a relay and the successfully received time of the packet at the destination in number of time slots (number of channel uses). 
A single packet is transmitted from the source at each time slot and its size is determined by the selected $\SR$ channel gain according to adaptive rate transmission.
Therefore, a delay of one means that a packet is stored at the relay in a certain time slot and all information bits contained in the packet are successfully forwarded to the destination in the next time slot. \Sumin{As stated in Section~\ref{sec:performance_evaluation}, we also consider 10000 packet transmissions from the source for the delay performance in following.}

\begin{figure}[tp]
\centering
\includegraphics[width=8.6cm]{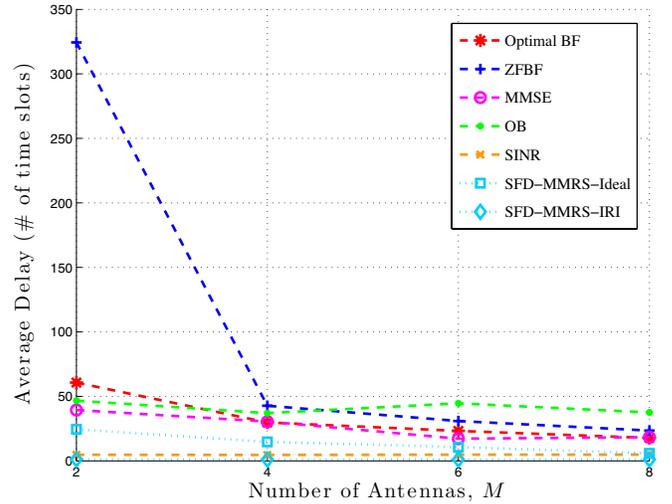}
\vspace{-3mm}
\caption{Average delay for varying number of antennas ($K=2$, $\mathrm{SNR}=20$ dB, $B_{max}\rightarrow\infty$, $\sigma_{\mathcal{S}\mathcal{R}_i}^2=\sigma_{\mathcal{R}_j\mathcal{D}}^2=\sigma_{\mathcal{R}_j\mathcal{R}_i}^2=0~\mathrm{dB},\forall i,j\in\mathcal{K}$)}
\label{fig:delay_varying_M}
\vspace{-5mm}
\end{figure}

In Fig.~\ref{fig:delay_varying_M}, we show the average delay for varying number of antennas when $K=2$, $\mathrm{SNR}=20$ dB, $B_{max}\rightarrow\infty$, and $\sigma_{\mathcal{S}\mathcal{R}_i}^2=\sigma_{\mathcal{R}_j\mathcal{D}}^2=\sigma_{\mathcal{R}_j\mathcal{R}_i}^2=0~\mathrm{dB},\forall i,j\in\mathcal{K}$.
Except for the proposed OB-based RS scheme, the average delays of all the schemes decrease as the number of antennas increases since the effective channel gain increases with the number of antennas. In the proposed OB-based RS scheme, the average delay is varying with the values slightly less than 50 time slots because its effective channel gain is same as the single antenna case.
The ideal SFD-MMRS scheme has moderate average delay and the non-ideal SFD-MMRS scheme and the SINR-based RS scheme have very short delays approaching one. The reason is that both schemes have the bottleneck in $\SR$ link due to uncoordinated IRI under i.i.d. channel condition and thus the size of packets transmitted from the source to a relay is much smaller than average $\RD$ link rate. 
On the contrary, the proposed optimal BF-based, ZFBF-based, and MMSE-based RS schemes have longer delays than that of the ideal SFD-MMRS scheme. In particular, the proposed ZFBF-based RS scheme has a significant delay when $M=2$ since its bottleneck is the $\RD$ link while the bottleneck of the MMSE-based RS scheme is the $\SR$ link in average sense. After all, the average packet size from the source in the ZFBF-based RS scheme is larger but its average $\RD$ link rate is smaller than those of the MMSE-based RS scheme.
However, the ZFBF-based RS scheme recovers the loss in the effective SNR at the destination and thus the average delay is significantly reduced when $M\geq 4$.
Similarly, in the optimal BF-based RS scheme, both average $\SR$ and $\RD$ link rates are more balanced than those in the MMSE-based RS scheme. 
This results in a larger average $\SR$ link rate and a smaller average $\RD$ link rate for the optimal BF-based RS scheme, compared to the MMSE-based RS scheme.
Accordingly, the average delay of the optimal BF-based RS scheme becomes worse than that of the MMSE-based RS scheme, since its average packet size is larger but its average $\RD$ link rate is smaller than those of the MMSE-based RS scheme.

\begin{figure}[tp]
\centering
\includegraphics[width=8.6cm]{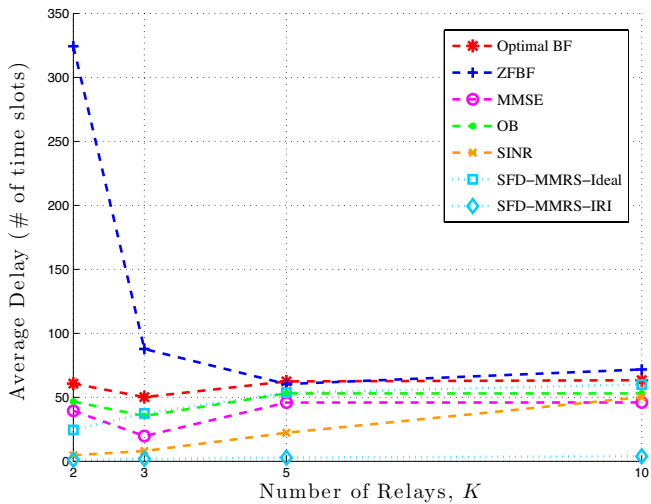}
\vspace{-3mm}
\caption{Average delay for varying number of relays ($M=2$, $\mathrm{SNR}=20$ dB, $B_{max}\rightarrow\infty$, $\sigma_{\mathcal{S}\mathcal{R}_i}^2=\sigma_{\mathcal{R}_j\mathcal{D}}^2=\sigma_{\mathcal{R}_j\mathcal{R}_i}^2=0~\mathrm{dB},\forall i,j\in\mathcal{K}$)}
\label{fig:delay_varying_K}
\vspace{-5mm}
\end{figure}

Fig.~\ref{fig:delay_varying_K} shows the average delay for varying number of relays in the same setup. 
The basic trend of the ideal SFD-MMRS scheme increases as the number of relays increases since the more relays exist, each relay is less likely to be selected.
The SINR-based RS scheme follows the similar trend while the non-ideal SFD-MMRS scheme has still very short delays approaching one. This is because the SINR-based RS scheme can provide much larger average link rates as the number of relays increases as shown in Fig.~\ref{fig:C_varying_K}.
The other proposed schemes optimizing beamformers have a different trend where the average delay decreases until $K=3$ ($K=5$ in the ZFBF-based RS scheme) and increases again as the number of relays increases further.
When $K=2$ and $M=2$, the minimum network setup, the degrees of freedom optimizing transmit and/or receive beamformers is too restricted but one additional relay gives an additional degree of freedom improving link rates. Hence, the average delays of the proposed schemes are rather decreased when $K=3$, but increased again after then, since the additional gain is almost saturated after a certain moment compared to the reduced selection opportunity per relay.
\Sumin{From Figs.~\ref{fig:delay_varying_M} and \ref{fig:delay_varying_K}, since the proposed ZFBF-based RS scheme requires too much delay at the minimum network setup ($K=2, M=2$), the MMSE-based RS scheme is more desirable for this setup in the perspective of the average delay performance.}

\subsection{Behavior of the Optimal Weight Factors}\label{subsec:opt_alpha}

\begin{figure}[tp]
\centering
\subfigure[]{
\includegraphics[width=8.6cm]{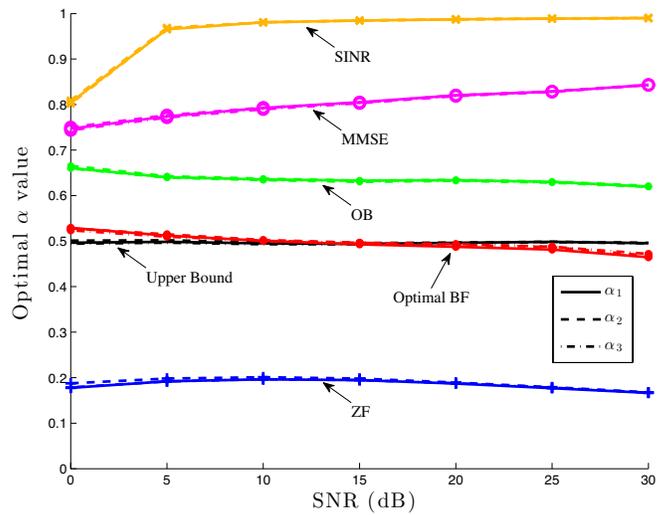}}
\subfigure[]{
\includegraphics[width=8.6cm]{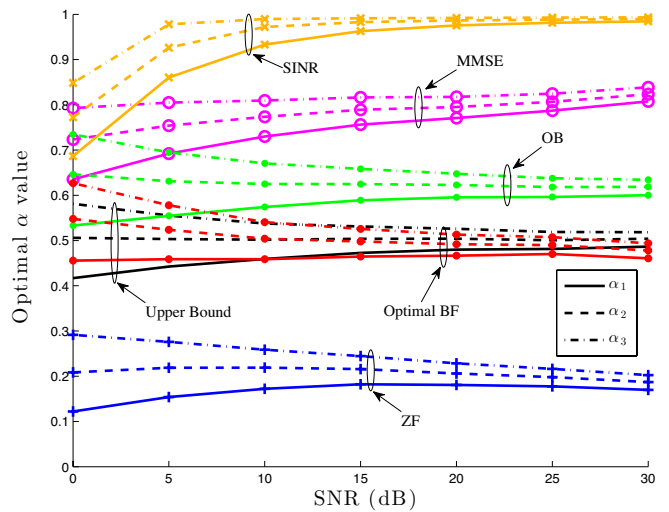}}
\caption{Optimal $\alpha$ value after a training phase ($K=3$, $M=2$) (a) i.i.d. channel case ($\sigma_{\mathcal{S}\mathcal{R}_i}^2=\sigma_{\mathcal{R}_j\mathcal{D}}^2=\sigma_{\mathcal{R}_j\mathcal{R}_i}^2=0~\mathrm{dB},\forall i,j\in\mathcal{K}$) (b) non-i.i.d. channel case ($[\sigma_{\mathcal{S}\mathcal{R}_1}^2,\sigma_{\mathcal{S}\mathcal{R}_2}^2,\sigma_{\mathcal{S}\mathcal{R}_3}^2]=[1,0,-1]~\mathrm{dB}$, $[\sigma_{\mathcal{R}_1\mathcal{D}}^2,\sigma_{\mathcal{R}_2\mathcal{D}}^2,\sigma_{\mathcal{R}_3\mathcal{D}}^2]=[-1,0,1]~\mathrm{dB}$, $\sigma_{\mathcal{R}_1\mathcal{R}_2}^2=\sigma_{\mathcal{R}_2\mathcal{R}_1}^2=\sigma_{\mathcal{R}_2\mathcal{R}_3}^2=\sigma_{\mathcal{R}_3\mathcal{R}_2}^2=0~\mathrm{dB},\sigma_{\mathcal{R}_1\mathcal{R}_3}^2=\sigma_{\mathcal{R}_3\mathcal{R}_1}^2=-1~\mathrm{dB}$)}
\label{fig:opt_a_K3_M2}
\vspace{-5mm}
\end{figure}

In this subsection, we investigate the behavior of the optimal $\alpha_k$ parameters for the proposed schemes in two different channel cases: \emph{i.i.d.} and \emph{non-i.i.d.} channel cases with three relays.
In simulations, to find the optimal $\alpha_k$ values, we employ the back-pressure algorithm approach using a pre-training phase as stated in \eqref{eq:backpressure}.
Fig.~\ref{fig:opt_a_K3_M2} shows the optimal $\alpha_k$ values obtained after the pre-training phase when $K=3$ and $M=2$.

Fig.~\ref{fig:opt_a_K3_M2}~(a) shows an i.i.d. channel case where all the links have an identical average channel quality as $0$ dB.
In the figure, all $\alpha_k$ values for each scheme are identical due to the i.i.d. channel condition.
In principle, if $\alpha_k$ is close to one, the $\SR$ link rate dominates the cost function and if $\alpha_k$ is close to zero, the $\RD$ link rate dominates the cost function in the relay pair selection.
The upper bound and the optimal BF-based RS scheme balance both link rates as $\alpha_k^{\star}\approx 0.5$ since the optimum is achieved when $\mathbb{E}[C_{\mathcal{SR}}(t)] = \mathbb{E}[C_{\mathcal{RD}}(t)]$.
In contrast, the optimal $\alpha_k$ values of the proposed MMSE-based and SINR-based RS schemes are close to one, while that of the proposed ZFBF-based RS scheme is close to zero. This is because the bottleneck of the MMSE-based and SINR-based RS schemes is the $\SR$ link while the bottleneck of the ZFBF-based RS scheme is the $\RD$ link.

Fig.~\ref{fig:opt_a_K3_M2}~(b) shows a non-i.i.d. channel case, the same setup in Fig.~\ref{fig:C_nid}. For all the schemes, $\alpha_k$'s have different values according to average channel quality and $\alpha_2$'s have almost the same values as those for the i.i.d. channel case since $\mathcal{R}_2$ has the same average channel quality as the i.i.d. channel case. In the figure, it holds that $\alpha_1 < \alpha_2 < \alpha_3$ because $\sigma_{\mathcal{SR}_1}^2 > \sigma_{\mathcal{R}_1\mathcal{D}}^2$, $\sigma_{\mathcal{SR}_2}^2 = \sigma_{\mathcal{R}_2\mathcal{D}}^2$, and $\sigma_{\mathcal{SR}_3}^2 < \sigma_{\mathcal{R}_3\mathcal{D}}^2$, and the optimal $\alpha_k$ parameter tends to be determined in order to balance both $\SR$ and $\RD$ links. 
As SNR increases, the differences among $\alpha_k$ values are reduced since imbalance effect among links is diminished at high SNR.
Similarly to the i.i.d. channel case, the upper bound and the optimal BF-based RS scheme are biased at $\alpha_k=0.5$, and the proposed MMSE-based and SINR-based RS schemes are biased to be close to one while the ZFBF-based RS scheme is biased to be close to zero.

\subsection{Effects of Finite Buffer Size}

\begin{figure}[tp]
\centering
\includegraphics[width=8.6cm]{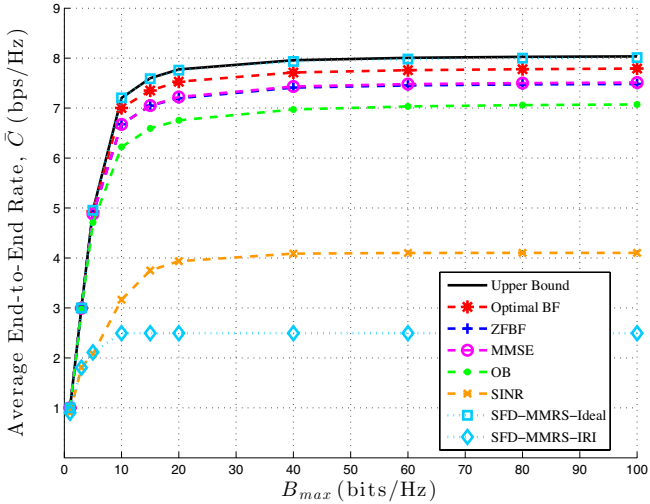}
\vspace{-3mm}
\caption{Average end-to-end rate with finite buffer size ($K=3$, $M=2$, $\mathrm{SNR}=20$ dB, $\sigma_{\mathcal{S}\mathcal{R}_i}^2=\sigma_{\mathcal{R}_j\mathcal{D}}^2=\sigma_{\mathcal{R}_j\mathcal{R}_i}^2=0~\mathrm{dB},\forall i,j\in\mathcal{K}$)}
\label{fig:C_avg_finite_buffer_size}
\end{figure}

In practice, the buffer size at relays is finite and thus it restricts the performance since a full-buffer relay cannot be selected as the receiving relay.
We investigate the performance in terms of average end-to-end rate and average delay according to the finite buffer size.
Fig.~\ref{fig:C_avg_finite_buffer_size} shows the average end-to-end rate for varying buffer size when $K=3$, $M=2$, $\mathrm{SNR}=20$ dB, and $\sigma_{\mathcal{S}\mathcal{R}_i}^2=\sigma_{\mathcal{R}_j\mathcal{D}}^2=\sigma_{\mathcal{R}_j\mathcal{R}_i}^2=0~\mathrm{dB},\forall i,j\in\mathcal{K}$.
All the schemes rapidly converge to their own performance upper limits with infinite buffer size as the buffer size increases.
Accordingly, a buffer size $B_{\max}\geq 50$ bits/Hz is sufficient to obtain the performance upper limits with infinite buffer size.
For instance, when the bandwidth is 10 MHz, the buffer size is required to be about 60 MB in order to achieve the average end-to-end rate with infinite buffer size.
This value is allowable at relays in the viewpoint of present memory size.

\begin{figure}[tp]
\centering
\includegraphics[width=8.6cm]{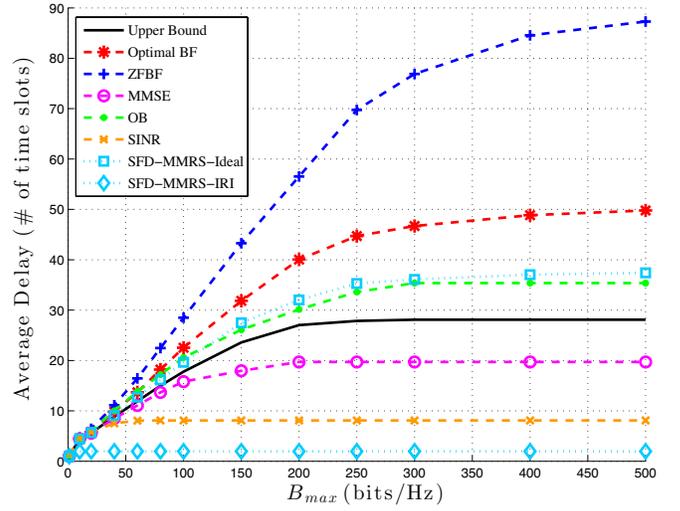}
\vspace{-3mm}
\caption{Average delay with finite buffer size ($K=3$, $M=2$, $\mathrm{SNR}=20$ dB, $\sigma_{\mathcal{S}\mathcal{R}_i}^2=\sigma_{\mathcal{R}_j\mathcal{D}}^2=\sigma_{\mathcal{R}_j\mathcal{R}_i}^2=0~\mathrm{dB},\forall i,j\in\mathcal{K}$)}
\label{fig:delay_finite_buffer_size}
\vspace{-5mm}
\end{figure}

Fig.~\ref{fig:delay_finite_buffer_size} shows the average delay performance with finite buffer size.
For all the schemes, the average delays converge to their own limits as the buffer size increases although the convergence speed is different between the schemes. 
Interestingly, the finite buffer size can be helpful to reduce the average delay without a loss in the average end-to-end rate. For example, even if the proposed ZFBF-based RS scheme has the longest delay, the average delay can be less than 15 time slots with achieving the average end-to-end rate limit if the buffer size is set to 50 bits/Hz.
Therefore, if the buffer size is set to an appropriate value, it is enough to achieve near-optimal average end-to-end rate under reasonable average delay.

\section{Conclusion} \label{sec:conclusion}

In this paper, we proposed virtual FD buffer-aided joint RS and BF schemes taking IRI into account in a buffer-aided multiple relays network, where each relay is equipped with multiple antennas. We first formulated a weighted sum-rate maximization based on instantaneous rates maximizing the average end-to-end rate. Based on the alternative objective function, we proposed various RS schemes based on optimal and suboptimal BF designs to cancel or suppress IRI taking a trade-off between computational complexity and performance into consideration. Through simulations, the proposed joint RS and BF schemes were evaluated in terms of the average end-to-end rate and average delay, compared to several conventional HD RS and SFD-MMRS schemes.
In numerical results, asymptotic trends were investigated with respect to the number of relays and the number of antennas at relays. The proposed joint RS and BF schemes recover the loss of multiplexing gain in the HD relaying even in the presence of IRI as the number of antennas and/or the number of relays increase. 
Although the complexities of the proposed joint RS and BF schemes are high due to global CSI required, they can be useful benchmarks in system simulations.
In addition, the behavior of the optimal weight factor and the effects of finite buffer size were shown in various different network setups.
Basic trend in the optimal weight factor is moving toward to reduce the link rate gap between the bottleneck link and the other link.
Although the finite buffer size limits the average end-to-end rate, it can help to bound the average delay with achieving near-optimal average end-to-end rate if it is set to an appropriate value. 
For future studies, it is possible to extend to multiple source-destination pairs with multiple antennas, to apply for non-full queue traffic at the source, to consider imperfect CSI and BSI, to develop a low complexity and limited feedback RS scheme, and to apply for other applications such as cognitive radio and physical layer security.

\section*{Acknowledgement}

\Sumin{We would like to thank the anonymous reviewers for constructive comments that helped us improve the quality of this paper including more practical concerns.}

\balance
\bibliographystyle{./style/IEEEtran_v111}
\bibliography{./style/IEEEabrv,./style/RefAbrv,Reference}

\end{document}